\documentclass{aa}  

\usepackage{graphicx}
\usepackage{txfonts}
\begin{document}

   \title{The evolution of the H$_2$O maser emission \\ in the accretion burst source G358.93$-$0.03\thanks{Table 4 is only available in electronic form
at the CDS via anonymous ftp to cdsarc.u-strasbg.fr (130.79.128.5)
or via http://cdsweb.u-strasbg.fr/cgi-bin/qcat?J/A+A/}}

   \author{O. S. Bayandina
          \inst{1}
          \and
          C. L. Brogan\inst{2}
          \and
          R. A. Burns\inst{3,4,5}
          \and 
          A. Caratti o Garatti\inst{6,7}
          \and
          J. O. Chibueze\inst{8,9}
          \and
          S. P. van den Heever\inst{10}
          \and
          S.~E. Kurtz\inst{11}
          \and 
          G. C. MacLeod\inst{10}
          \and
          L. Moscadelli\inst{1}
          \and
          A. M. Sobolev\inst{12}
          \and
          K. Sugiyama\inst{13,4}
          \and
          I. E. Val'tts\inst{14}
          \and
          Y. Yonekura\inst{15}
          }

             \institute{INAF – Osservatorio Astrofisico di Arcetri, Largo E. Fermi 5, 50125 Firenze, Italy\\
             \email{olga.bayandina@inaf.it}
             \and
             National Radio Astronomy Observatory, 520 Edgemont Road, Charlottesville, VA 22903, USA
             \and
             Joint Institute for VLBI ERIC, Oude Hoogeveensedijk 4, 7991 PD Dwingeloo, The Netherlands
             \and
             Mizusawa VLBI Observatory, National Astronomical Observatory of Japan, 2-21-1 Osawa, Mitaka, Tokyo 181-8588, Japan
             \and
             Korea Astronomy and Space Science Institute, 776 Daedeokdae-ro, Yuseong-gu, Daejeon 34055, Republic of Korea
             \and
             INAF - Osservatorio Astronomico di Capodimonte, via Moiariello 16, 80131, Napoli, Italy
             \and
             Dublin Institute for Advanced Studies, School of Cosmic Physics, Astronomy \& Astrophysics Section, 31 Fitzwilliam Place, Dublin 2, Ireland
             \and
             Space Research Unit, Physics Department, North-West University, Potchefstroom, 2520, South Africa
             \and
             Department of Physics and Astronomy, University of Nigeria, Carver Building, 1 University Road, Nsukka, 410001, Nigeria
             \and
             SARAO, Hartebeesthoek Radio Astronomy Observatory, PO Box 443, Krugersdorp, 1741, South Africa
             \and
             Instituto de Radioastronom{\'\i}a y Astrof{\'\i}sica, Universidad Nacional Aut\'onoma de M\'exico, Antig. Carr. a Patzcuaro 8701, Morelia, 58089, Mexico
             \and
             Ural Federal University, 51 Lenin Str., 620051 Ekaterinburg, Russia
             \and
             National Astronomical Research Institute of Thailand (Public Organization), 260 Moo 4, T. Donkaew, A. Maerim, Chiang Mai, 50180, Thailand
             \and
             Astro Space Center, P.N. Lebedev Physical Institute of RAS, 84/32 Profsoyuznaya st., Moscow, 117997, Russia
             \and
             Center for Astronomy, Ibaraki University, 2-1-1 Bunkyo, Mito, Ibaraki 310-8512, Japan
             }

   \date{Received 23/05/2022; accepted 15/06/2022}

 
  \abstract
   {The massive young stellar object (MYSO) \object{G358.93$-$0.03-MM1} showed an extraordinary near-infrared- to (sub-)millimetre-dark and far-infrared-loud accretion burst, which is closely associated with flares of several class II methanol maser transitions, and, later, a 22 GHz water maser flare.}
   {Water maser flares provide an invaluable insight into  ejection events associated with accretion bursts. Although the 
      short timescale of the 22 GHz water maser flare made it impossible to carry out a very long baseline interferometry observation, we could track it with the Karl
G. Jansky Very Large Array (VLA).}
   {The  evolution of the spatial structure of the 22 GHz water masers and their association with the continuum sources in the region is studied with the VLA during two epochs, pre- and post-H$_2$O maser flare.}
   {A drastic change in the distribution of the water masers is revealed: in contrast to the four maser groups detected during epoch I, only two newly formed clusters are detected during epoch II. The 22 GHz water masers associated with the bursting source MM1 changed in morphology and emission velocity extent.}
   {Clear evidence of the influence of the accretion burst on the ejection from \object{G358.93$-$0.03-MM1} is presented. The accretion event has also potentially affected a region with a radius of $\sim$2$\arcsec$ ($\sim$13~500~AU at 6.75 kpc), suppressing water masers associated with other point sources in this region.
  }


   \maketitle
%

\section{Introduction} \label{sec:intro}

  Disk-mediated accretion accompanied by episodic accretion bursts is thought to be a common mechanism of both low- and high-mass star formation \citep[e.g.][]{Crimier2010, Meyer19}. However, in the case of high-mass protostars, this phenomenon is poorly studied as it has only recently been discovered \citep{Caratti17}.
  Little observational evidence exists, and only four accretion bursts in massive young stellar objects (MYSOs)  have been detected and analysed so far \citep{Caratti17, Hunter17, Proven19, MacLeod2021, stecklum21}. Recent studies show that maser monitoring can identify and trace high-mass young stellar objects (YSOs) at the very important, yet very short, phase of accretion bursts \citep[e.g.][]{Szymczak18, MacLeod18, MacLeod19,  Brogan19, Breen19}.

\object{G358.93$-$0.03-MM1}, located at the near kinematic distance of $6.75^{+0.37}_{-0.68}$~kpc \citep{Reid2014}, showed an extraordinary near-infrared-  to (sub-)millimetre-dark and far-infrared-loud MYSO accretion burst in 2019 \citep{stecklum21}. The \object{G358.93$-$0.03} region is densely populated, and \cite{Brogan19} resolved the protocluster into eight (sub-)millimetre continuum sources designated as MM1-MM8 in order of decreasing right ascension (RA). 
The continuum source \object{G358.93$-$0.03-MM1} shows characteristics of an extremely young massive object \citep{stecklum21}. 
The circumstellar disk around MM1 is found to be unusually low-mass for an MYSO, and the accretion burst in the source is the least energetic one out of the limited sample of such events in MYSOs \citep{stecklum21}. 

The study of the maser emission during the burst performed by the Maser Monitoring Organization (M2O\footnote{\url{https://www.masermonitoring.org/}}; a global community for maser-driven astronomy; \citealt{Burns22}) has provided the most complete picture of an accretion burst in MYSOs to date \citep{MacLeod19,  Brogan19, Breen19, Burns20, stecklum21, Bayandina2022}.
The MM1 burst was accompanied with flares of maser emission \citep[e.g.][]{Sugiyama19, MacLeod19, Brogan19, Breen19, Bayandina2022}. Among the detected masers were rare and even previously undiscovered methanol maser transitions, some of which were unpredicted \citep{MacLeod19, Breen19, Brogan19}. 
We note that the maser flare in the source was detected by the M2O first  \citep{Sugiyama19, MacLeod19,  Brogan19, Breen19} and only later was the study of the infrared data able to confirm the accretion burst \citep{stecklum21}.

Interferometric observations, conducted after the methanol maser flare, revealed the presence of an accretion disk  and were able to follow the propagation of a heatwave across it  \citep{Burns20,Chen20b,Bayandina2022}. 
The next important step was to study the spatial structure of the water maser. Accretion and  ejection are known to be closely associated - brightening outflow cavities \citep{Caratti17}, flares of H$_2$O masers likely tracing jets \citep{Chibueze2021}, and radio jet bursts \citep{Cesaroni2018} have been reported after accretion bursts.

Due to the short duration of the water maser flare in \object{G358.93-0.03}, no very long baseline interferometry (VLBI) observations were made, and the low flux density of the H$_2$O maser outside of the flare made it difficult to image with VLBI. However, the faint water maser emission is readily detectable by the Karl
G. Jansky Very Large Array (VLA). With the high sensitivity and moderate resolution of the VLA, we were able to study the spatial structure of the H$_2$O masers  before and after the maser flares. 
Additionally, with the VLA we were able to study the centimetre-continuum sources in the region.

\begin{table*}
\caption{Observation parameters: Continuum.}             
\label{tab:obsc}      
\centering                          
\begin{tabular}{cccccccccc}        
\hline\hline                 
Band\tablefootmark{a} & Freq. & Epoch & Int. Time & Synth. Beam & PA & 1$\sigma$~rms\tablefootmark{b} & \multicolumn{2}{c}{Detect.} \\  
\cline{8-9}
 & (GHz) &  & (min) & (arcsec)&  ($^{\circ}$) &  ($\mu$Jy/beam) & MM1\tablefootmark{c} & MM3\tablefootmark{c} \\
\hline                        
C$\tablefootmark{1}$  & 6.0 & I & 14 & 4.66~$\times$~0.77 & +9.02 
& 23 & N & Y\\
   &  & II & 16 & 2.61~$\times$~1.05 & $-$0.07 
   & 23 & N &Y \\
   \hline
Ku$\tablefootmark{2}$ &  15.0 & I& 16 & 1.84~$\times$~0.46 & $+$3.29 
& 11 &  N &Y \\
   &   & II & 16 & 1.12~$\times$~0.45 & $-$6.32 
  & 13 & N & Y\\
  \hline
K$\tablefootmark{3}$  &  20.0 & I  & 32 & 1.28~$\times$~0.30 & +0.28 
& 16 & Y & Y \\
   &   & II & 19 & 0.78~$\times$~0.32 & $-$9.98 
   & 15 & N & N \\
\hline                                   
\end{tabular}
\tablefoot{
\tablefoottext{a}{The following maser lines were detected in the continuum windows (the listed lines were flagged during the data processing to avoid false detections in the continuum images): 
 (1) 6.18, 6.67, 7.68, and 7.83 GHz; (2) 14.30 GHz; 
 (3) 20.97 and 20.35 (epoch I only) GHz.} \\
\tablefoottext{b}{\footnotesize{The detection threshold is set to the 3$\sigma$ level.}} \\
\tablefoottext{c}{\footnotesize{Hereinafter we adopt the source naming from \cite{Brogan19}.}}
}
\end{table*}

\section{Observations and data reduction \label{sec:obs}}

\subsection{VLA observations}

The observation of \object{G358.93$-$0.03} was conducted with the Karl G. Jansky VLA in two sessions: epoch I on February 25, 2019 (project code 19A-448, C$\,\to\,$B configuration), and epoch II on  June 4, 2019 (19A-476, A configuration). The H$_2$O maser flare in \object{G358.93$-$0.03} took place in April 2019 (Fig.~\ref{fig:timeline}), and thus the VLA epoch I observations of February 2019 are considered as the H$_2$O maser pre-flare epoch and the epoch II observations of June 2019  as the post-flare epoch.

The general technical details of the observations are described in \cite{Bayandina2022}. Here we mention only the parameters of the water maser study. 

The observation parameters of continuum and water maser data are presented in Tables \ref{tab:obsc} and \ref{tab:obsspl}, respectively.
To observe continuum emission, we used 31 spectral windows of 128 1 MHz channels in the C, Ku, and K bands. The maser line was observed in a 4 MHz spectral window of 1024 channels. 
 Detection of continuum and maser emission in each band and epoch is marked with `Y'; non-detection is marked with `N'. 

Data reduction was performed as described in \cite{Bayandina2022}. Here we repeat  two remarks detailed in \cite{Bayandina2022}: (1) the observations were made during a multi-frequency maser flare that led to a leakage of a few maser lines into continuum spectral windows (see the notes in Table \ref{tab:obsc}); (2) a positional shift was introduced into the maser data to compensate for a discrepancy in the absolute coordinates of the maser and continuum data obtained with different facilities and beams; for example, the maser data obtained with the VLA in the present work is compared with the Atacama Large Millimeter/submillimeter Array (ALMA) data from \cite{Brogan19}.

\begin{table*}
\caption{Observation parameters: Spectral line.}
\label{tab:obsspl}
\centering  
\begin{tabular}{cccccccccc}
\hline\hline   
Band & Transition & Freq. & Epoch & Int. Time &
Synth. Beam & PA & Spec. res. & $1\sigma$~rms & Detection\\ 
 &  & (MHz) & & (min) &
(arcsec) & ($^\circ$) & (km~s$^{-1}$) & (mJy/beam) & \\
\hline 
K & H$_2$O 6$_{1,6}\,\to\,5_{2,3}$ & 22 235.08 & I & 16 & 1.04~$\times$~0.21 & +0.64 & 0.05 & 14 & Y\\
    &  &   & II & 19 & 0.57~$\times$~0.24 & -8.35 & 0.11 & 17 & Y \\
\hline                                   
\end{tabular}
\end{table*}

\subsection{HartRAO monitoring}

The single-dish monitoring results reported here were made using the 26~m
telescope at the Hartebeesthoek Radio Astronomy Observatory
(HartRAO)\footnote{See http://www.hartrao.ac.za/spectra/ for further
information.}.  The 1.3 cm receiver
is a dual, left and right circularly polarised, cryogenically
cooled receiver. It was calibrated relative to Hydra~A, 3C123, and Jupiter
assuming the flux scale of \citet{Ott94}. Observations were recorded with
a 1024-channel (per polarisation), 1.0~MHz bandwidth spectrometer.
Frequency switching was employed as were half-power beamwidth pointing
correction observations for all epochs of observation. The total velocity
extent of each observation is 54 km~s$^{-1}$, and the resolution is 0.105 km~s$^{-1}$. The
observing frequency is 22.235120 MHz, and this is corrected for the local
standard of rest (LSR) velocity $v_{LSR} = -17~$km~s$^{-1}$. Each epoch of
observation comprises two six-minute observations with a
sensitivity of $\sim$0.1~K or $\sim$1~Jy per polarisation. The
beamwidth is 2.2$\arcmin$. More information regarding this receiver and the
observing method employed is given in \citet{MacLeod18}. Monitoring
began on January 20,  2019, and observations were made every 10 to 20 days, subject to other scheduled observing. 

\section{Results} \label{sec:results}
\subsection{Continuum emission}

To provide context for the discussion of the water maser spatial distribution, we first present the continuum sources detected in the region.

\begin{figure*}
\centering
\begin{tabular}{cc}
  \includegraphics[width=75mm]{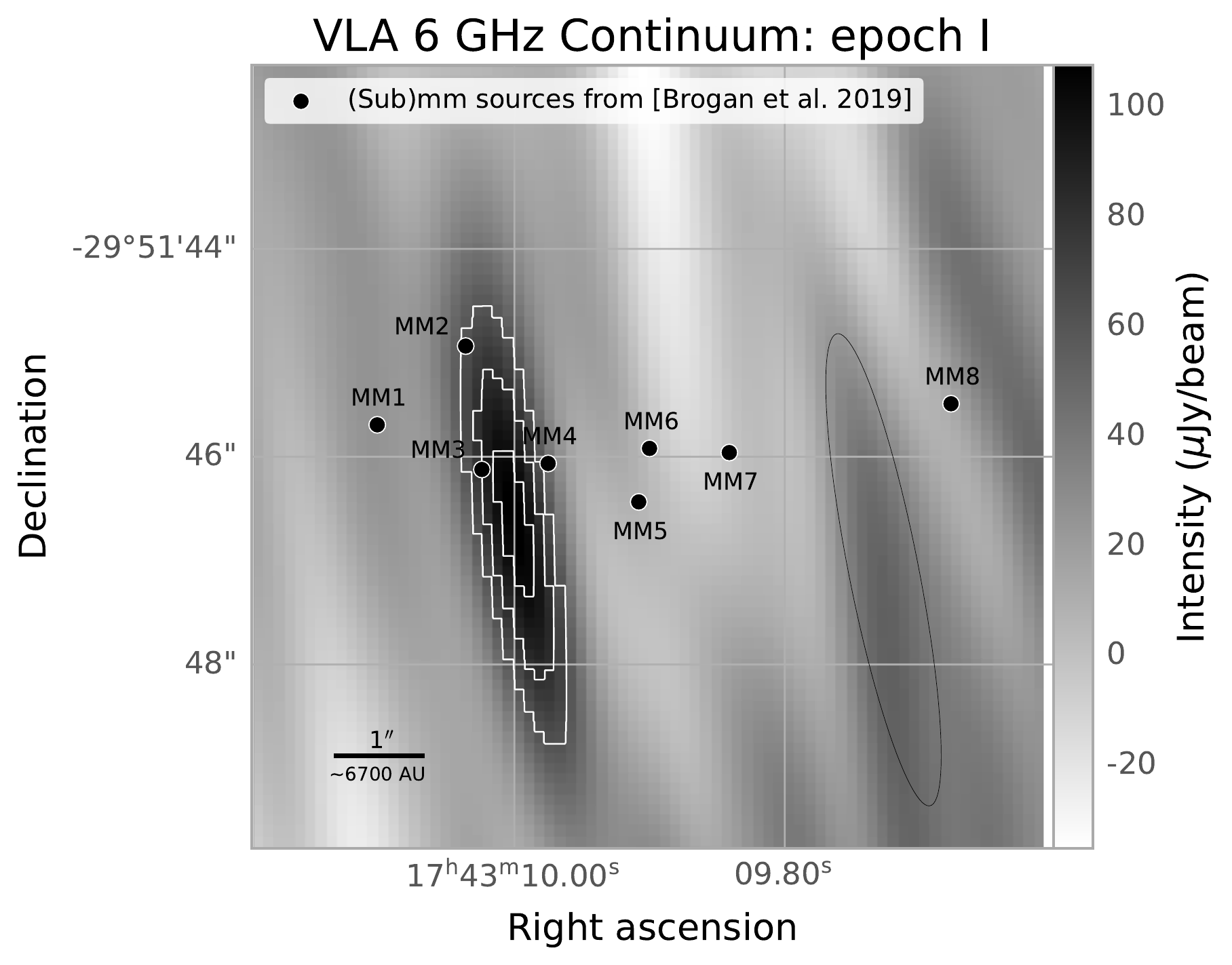} &   
  \includegraphics[width=75mm]{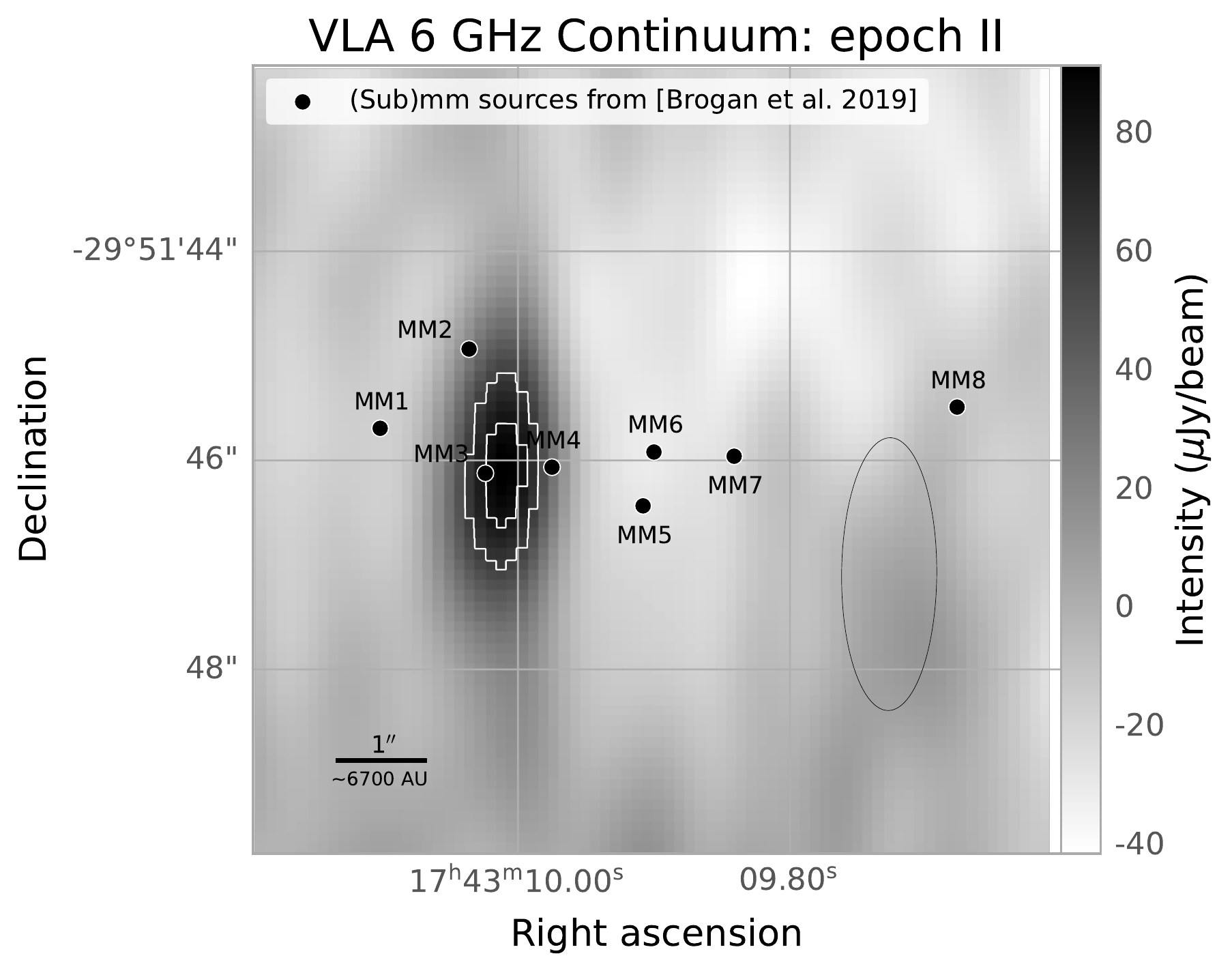} \\
(a)  & (b)  \\[6pt]
 \includegraphics[width=75mm]{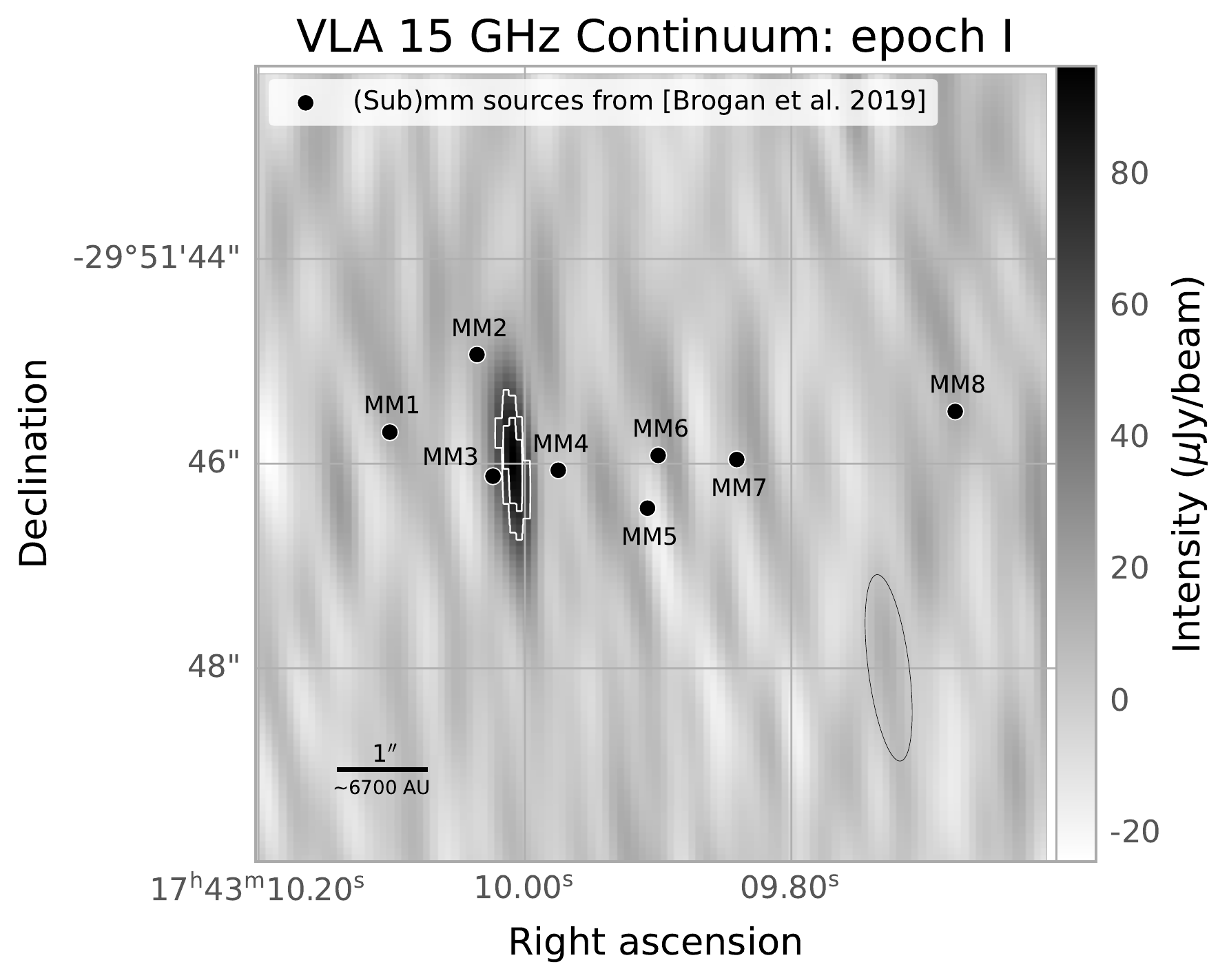} &   
 \includegraphics[width=75mm]{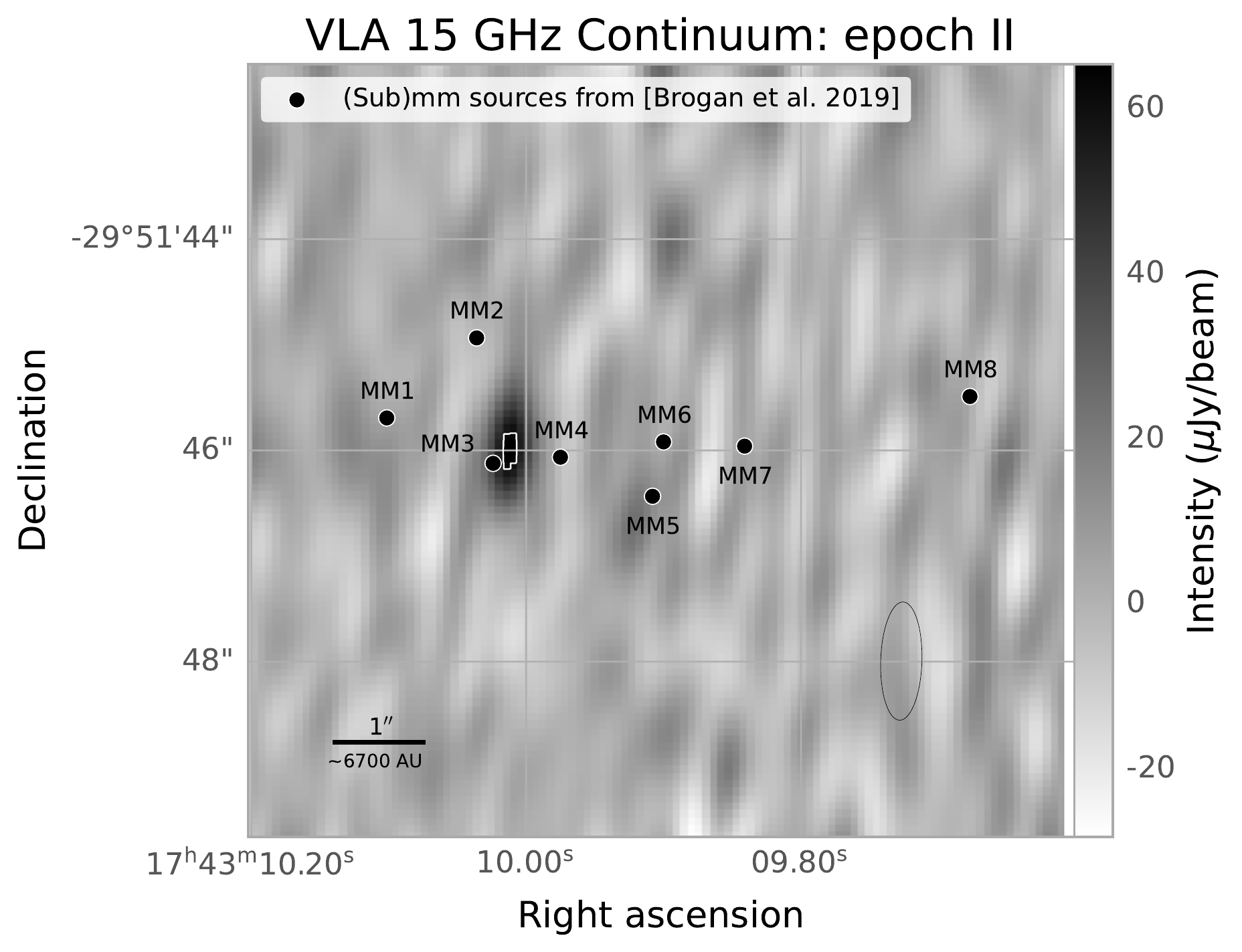} \\
(c)  & (d)  \\[6pt]
\includegraphics[width=75mm]{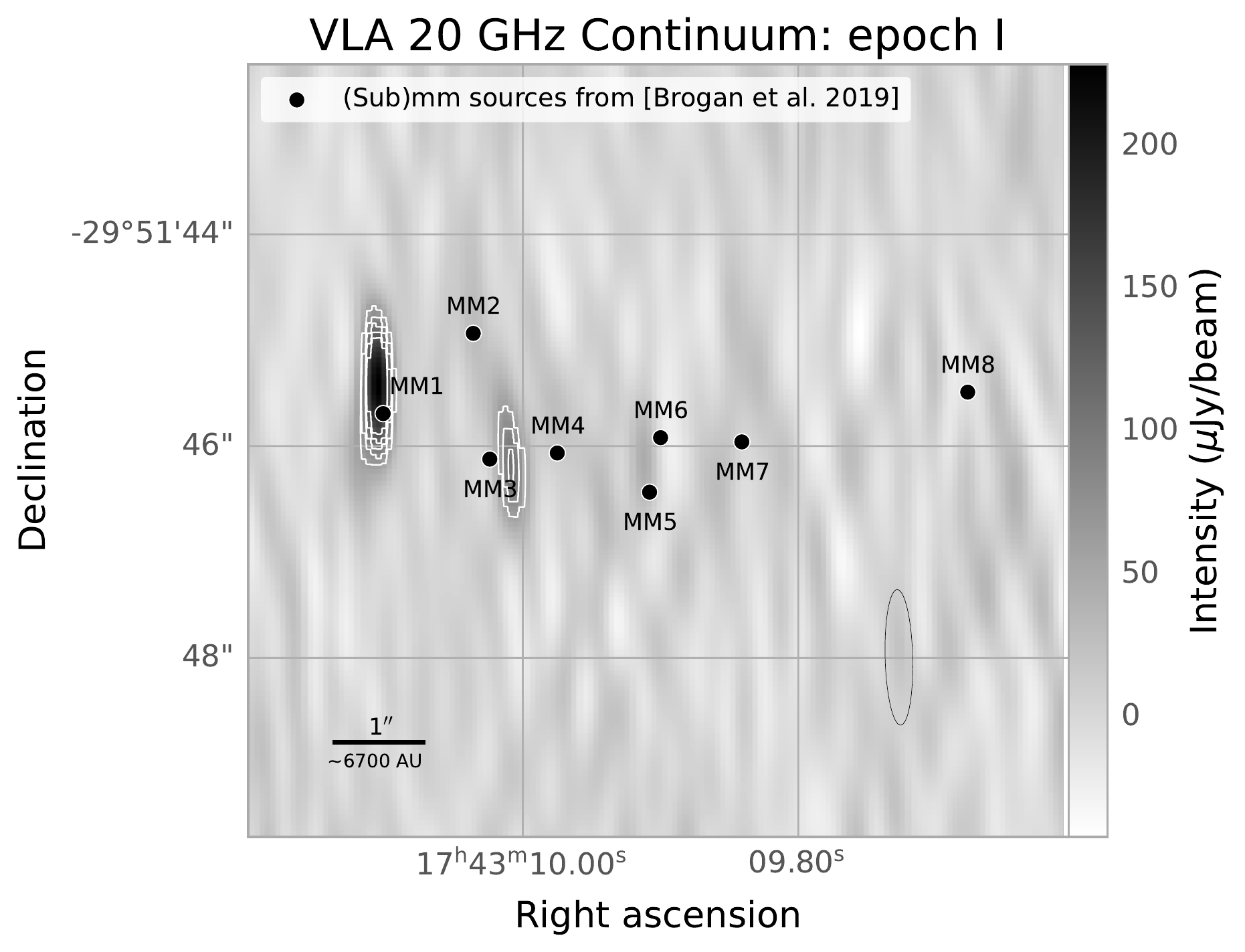} &   
\includegraphics[width=75mm]{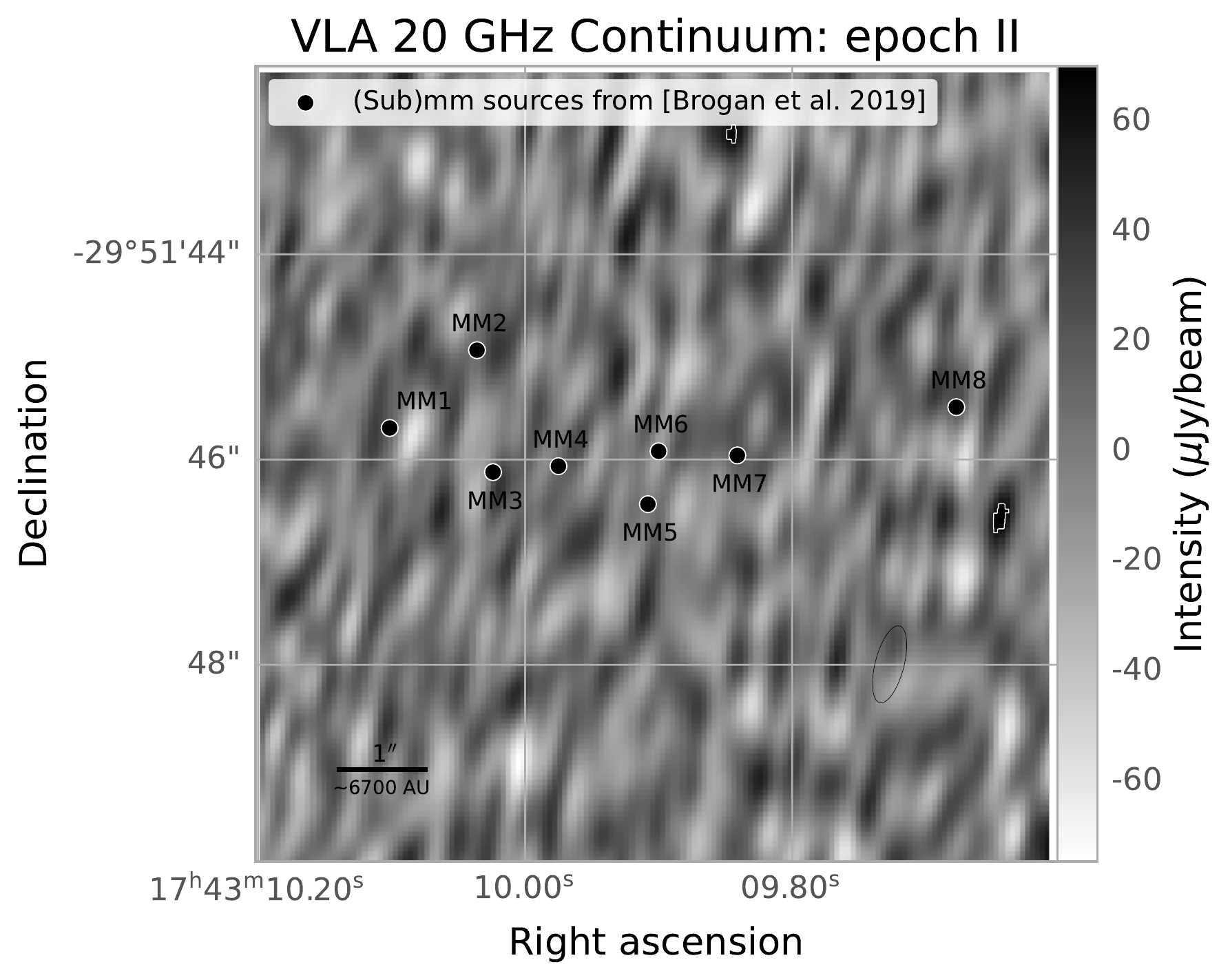} \\
(e)  & (f)  \\[6pt]
\end{tabular}
\caption{VLA continuum images of \object{G358.93$-$0.03} at 6, 15, and 20 GHz during epochs I (left panels) and II (right panels). The contour levels are
[6, 8, 10] $\times$ 10~$\mu$Jy/beam. 
The black dots mark the positions of the (sub-)millimetre sources from \cite{Brogan19}. The synthesized VLA beam size of each image is shown with the black ellipses in the lower-right corner of the panels.   \label{fig:cont}}
\end{figure*}

Our VLA observations towards \object{G358.93$-$0.03} detected two continuum sources associated with the hot cores MM1 and MM3 discovered in \cite{Brogan19}. The obtained VLA images are shown in Fig. \ref{fig:cont}, and a summary of the detected source parameters is presented in Table \ref{tab:cont}. 

MM1 is the brightest millimetre-continuum source, the most line-rich hot core in the region, and the source of the accretion burst  \citep{Brogan19}. As described in \cite{Bayandina2022}, MM1 was detected only in the K band during the first epoch of the VLA observations. 

\begin{figure}[ht]
\centering
\includegraphics[width=0.45\textwidth]{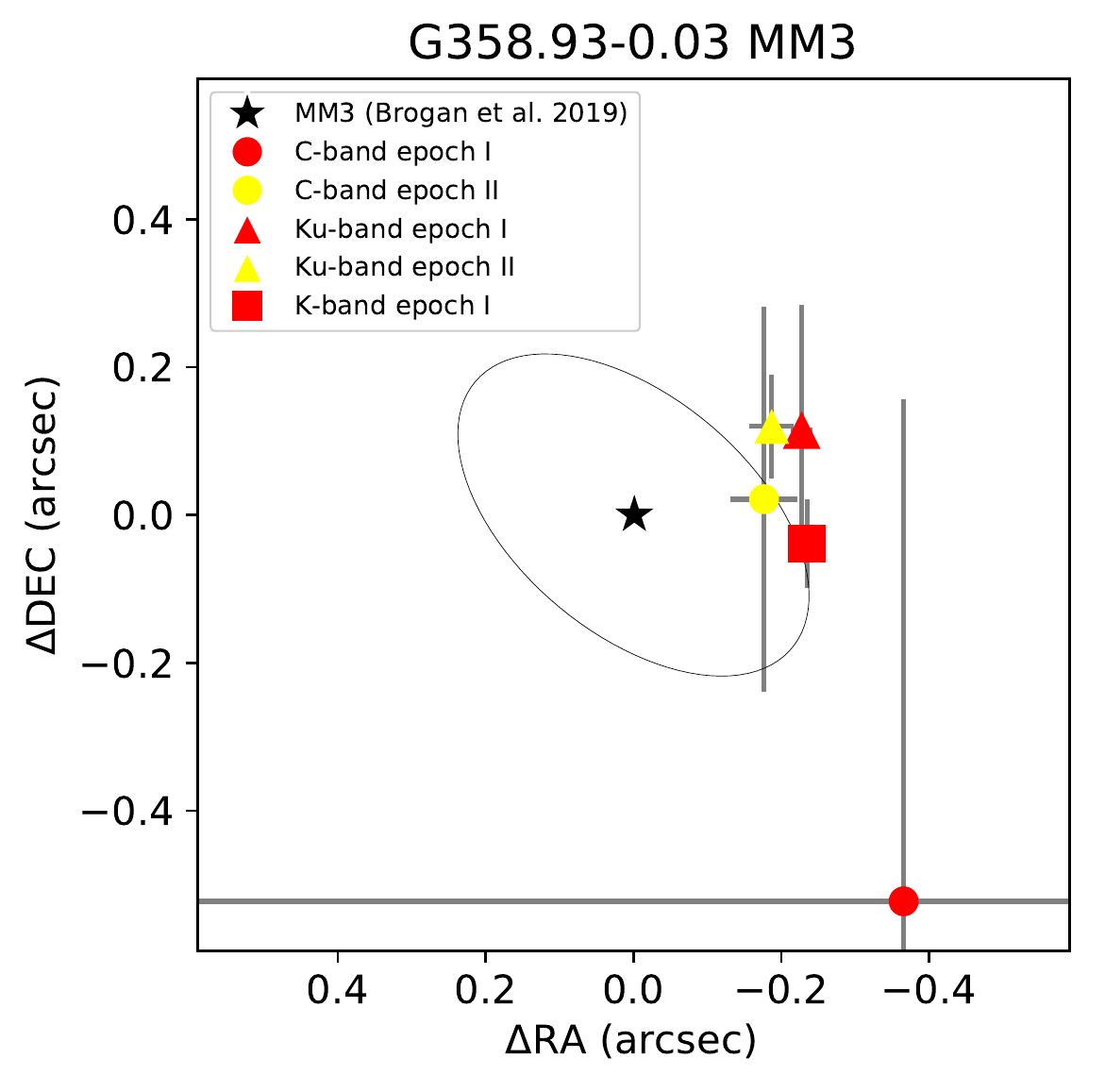}
\caption{Centimetre-continuum (this work) and millimetre-continuum \citep{Brogan19} peak position 
of MM3. The C- (circle), Ku- (triangle), and K-band (square) peaks detected with the VLA during epoch I are shown in red, and during epoch II in yellow. The ellipse represents the size estimation of MM3 obtained in \cite{Brogan19}. The grey error bars indicate the position fitting errors from Table~\ref{tab:cont}. 
\label{fig:position_check}}
\end{figure}

The continuum source detected in all three frequency bands with the VLA is associated with the ALMA hot core MM3 \citep{Brogan19}. The C-band epoch I detection is marginal, as it barely meets the 3$\sigma$ threshold. The ALMA and VLA positions of MM3 differ by $\sim$0.3$\arcsec$ (Fig. \ref{fig:position_check}), which is similar to the absolute positional uncertainty reported in \cite{Brogan19}.
The VLA and ALMA \citep{Brogan19} studies of MM3 flux density 
were done in the period of high activity in the region, which means we cannot directly compare the millimetre and centimetre fluxes.

No continuum emission above the 3$\sigma$ noise level is detected during VLA epoch II in the K band.

\begin{table*}
\caption{VLA continuum emission peak parameters.}
\label{tab:cont}
\centering  
\begin{tabular}{cccccccc}
\hline\hline   
Association &
Band & 
Epoch &
RA(J2000) & Dec.(J2000) & Integrated flux & Peak flux & SNR \\
 & &  & ($^h$~$^m$~$^s$) & ($^\circ$~$\arcmin$~$\arcsec$) & 
($\mu$Jy) & ($\mu$Jy/beam) & \\
\hline 
MM3  & C & I &  17:43:10.005(0.113)\tablefootmark{a} & $-$29:51:45.90(0.68)\tablefootmark{a} & 263(76) & 170(26) & 3\\
  & & II &  17:43:10.007(0.003) & $-$29:51.46:35(0.26) & 83(26) & 98(14) & 4\\ \cline{2-8}
& Ku  & I &   17:43:10.004(0.001) &  $-$29:51:46.27(0.17) & 132(27) & 109(11) & 11\\ 
&   & II &  17:43:10.007(0.002) & $-$29:51:46.30(0.07) & 113(19) & 84(9) & 8\\ \cline{2-8}
& K & I & 17:43:10.009(0.0004)  &  $-$29:51:46.03(0.06) & 133(14) & 111(6) & 11\\ 
\hline                                   
\end{tabular}
\tablefoot{
\tablefoottext{a}{The positional uncertainties are statistical errors of fit.}}
\end{table*}

\subsection{Water masers}

\subsubsection{HartRAO monitoring}

\begin{figure}[ht]
\centering
\includegraphics[width=0.55\textwidth]{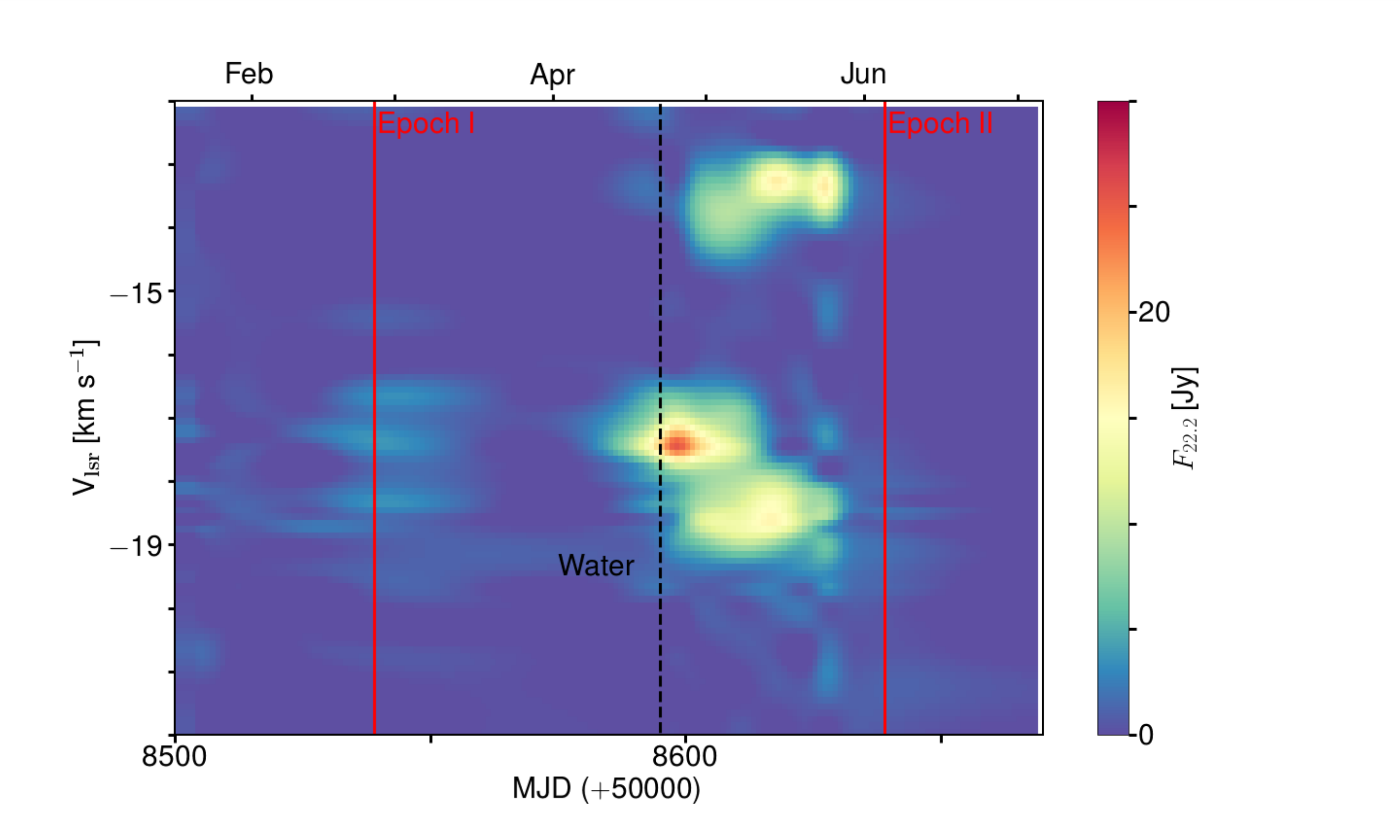}
\caption{Dynamic spectrum of the 22 GHz water maser emission in \object{G358.93$-$0.03} for the period January--July 2019. The black dotted line indicates the flare of the water maser. The red lines indicate the dates of the VLA epoch I and II observations (M2O data: the HartRAO 26 m telescope monitoring program).
\label{fig:timeline}}
\end{figure}

The flare of the water emission in \object{G358.93-0.03} was short-lived (each flare had a duration of about a day; see Fig.~\ref{fig:timeline}) and moderate (the maximum flux was $\sim$20~Jy). 
In contrast, the previously known accretion bursts in the MYSOs \object{NGC 6334I-MM1} and \object{S255 NIRS 3} showed long-lasting flares (timescales of a year to a few years)  of strong (kJy) H$_2$O masers \citep{MacLeod18, Hirota2021}.

The methanol masers in \object{G358.93-0.03-MM1} flared in February-March of 2019 \citep{MacLeod19,  Brogan19, Breen19}, but the water maser remained stable during this period, with  a  flux  density  of $\sim$1  Jy (see Fig. \ref{fig:timeline}). In April 2019, the single-dish telescopes participating in the M2O monitoring of \object{G358.93$-$0.03} detected a rapid and steady growth of the water maser flux density. By April 19, the spectral feature at V$_{LSR}$=$-$17.42 km s$^{-1}$ reached $\sim$26 Jy (Fig. \ref{fig:timeline}). Two more flares of the H$_2$O maser emission were noticed later. On May 14, 2019, the two spectral features at V$_{LSR}$=$-$13.31 km s$^{-1}$ and V$_{LSR}$=$-$18.58 km s$^{-1}$ flared simultaneously, reaching a flux density of 21 and 19 Jy, respectively. On May 23, 2019, the last case of enhanced activity of the water maser was detected when the spectral feature at V$_{LSR}$=$-$13.42 km s$^{-1}$ increased its flux to 22~Jy (the M2O monitoring; Fig. \ref{fig:timeline}).

\subsubsection{VLA}

\begin{figure*}[ht]
\centering
\includegraphics[width=0.85\textwidth]{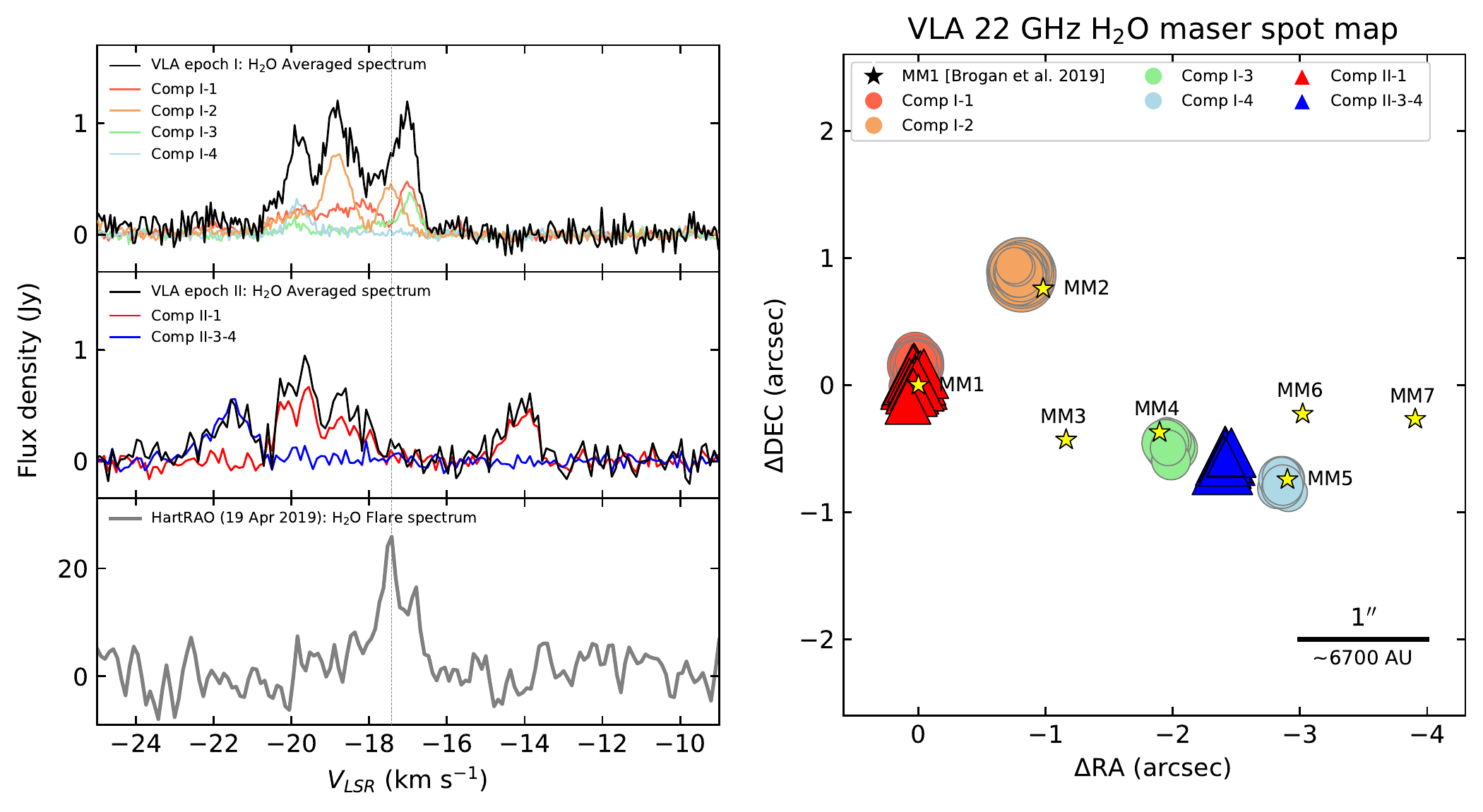}
\caption{{Comparison of the 22 GHz H$_2$O maser emission detected with the VLA during epochs I and II. \textit{Left panels:} Spectra of the H$_2$O maser components (marked by colour; see the legend) and total spectrum (black line) detected with the VLA during epoch I \textit{(top)} and epoch II \textit{(middle)}. The \textit{bottom} panel shows the maser flare spectrum obtained with the HartRAO 26 m telescope on April 19, 2019. The dashed line indicates the peak velocity of the flare spectrum. \textit{Right panel:} Combined H$_2$O maser spot map. The epoch I maser spots  are marked by light-coloured circles; the epoch II maser spots are marked  by bright-coloured triangles. Positional offsets are relative to MM1 \citep{Brogan19}. The position of the continuum emission detected with the VLA and associated with MM3 \citep{Brogan19} is marked by a black cross.} \label{fig:22compare}}
\end{figure*}

The H$_2$O maser spectra detected towards G358.93$-$0.03 during VLA epochs I and II differ significantly from each other (see the left panels of Fig.~\ref{fig:22compare}).
During VLA epoch I, the H$_2$O emission covered the velocity range from $\sim$$-$21 to $-$16 km~s$^{-1}$, and during epoch II from $\sim$$-$23 to $-13$ km~s$^{-1}$. During both epochs, three $\sim$1~Jy spectral features populate the spectra; however, their peak velocities are  different. The epoch I peaks are at V$_{LSR}$=$-$19.9, $-$18.8, and $-$17.0 km~s$^{-1}$, while the epoch II peaks are at V$_{LSR}$=$-$21.5, $-$19.6, and $-$13.8 km~s$^{-1}$. 
Notably, during the post-flare epoch (VLA epoch II) there is no emission at the velocity of the flaring spectral feature (Fig.~\ref{fig:22compare}).  

The position, velocity, and integrated and peak flux density of each of the detected 22 GHz water maser spots\footnote{In the following, we use the term `maser spots' to refer to maser emissions detected in a single velocity channel of a data cube.} are listed in Table~\ref{tab:T22GHZ}. 
The spot maps presented in the paper are based on  Table~\ref{tab:T22GHZ}. 
We note that the warning about the limitation of the interpretation of spatial structures visible on the VLA spot maps from \cite{Bayandina2022} applies to the water maser data as well.
In particular, even though the achieved signal-to-noise ratio allows us to fit the maser spot positions with sub-beamsize accuracy (Table~\ref{tab:T22GHZ}) and 
the structures seen in the water maser spot maps are much smaller than the synthesized VLA beam (Table \ref{tab:obsspl}), the spatial and velocity structure may be dominated by the brighter components.
The presented spot maps provide a general view of the spatial and velocity distribution of the water maser emission, but higher resolution observations are required to confirm the detected patterns.

\begin{figure*}[ht]
\centering
\begin{tabular}{cc}
  \includegraphics[width=80mm]{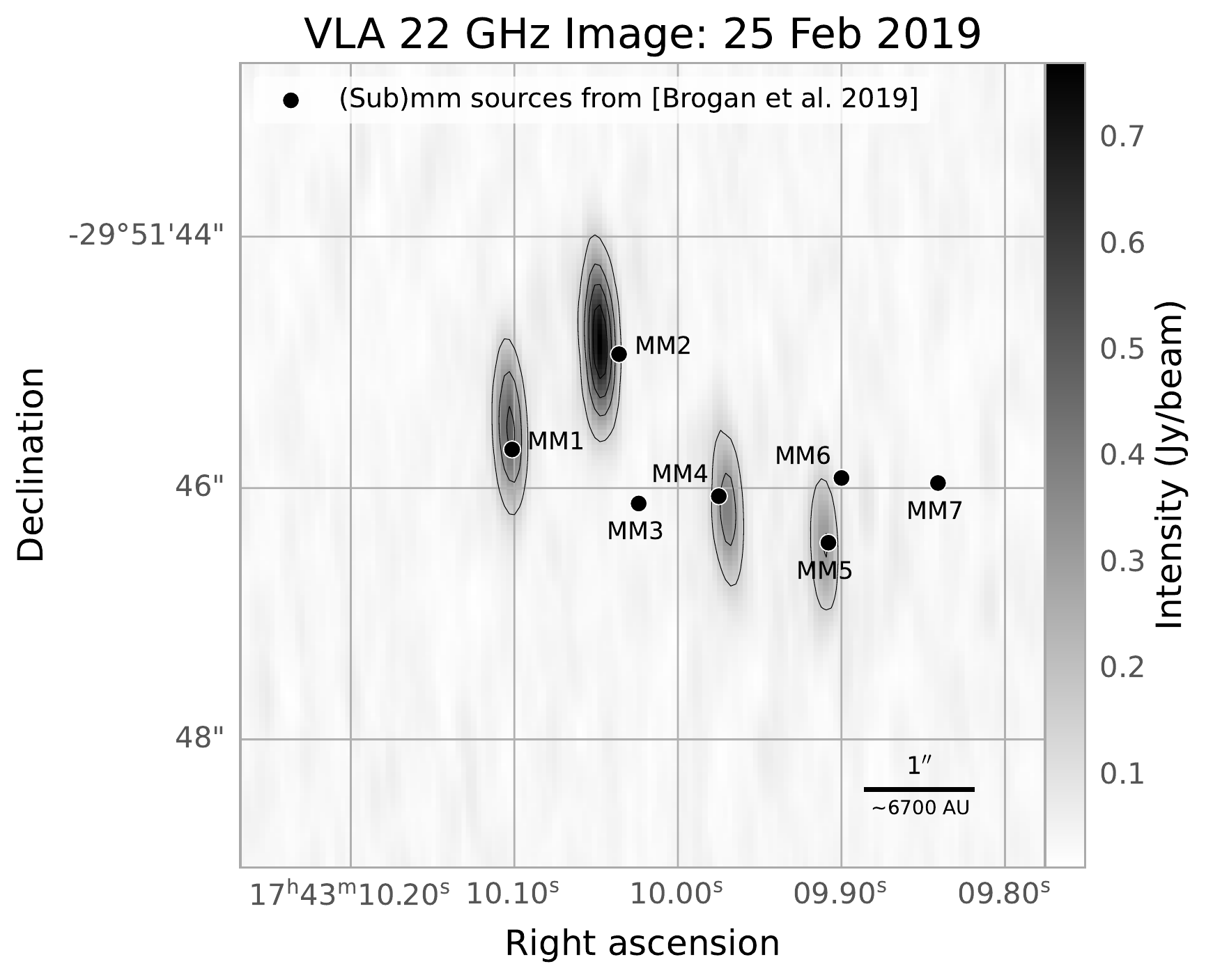} &   
  \includegraphics[width=80mm]{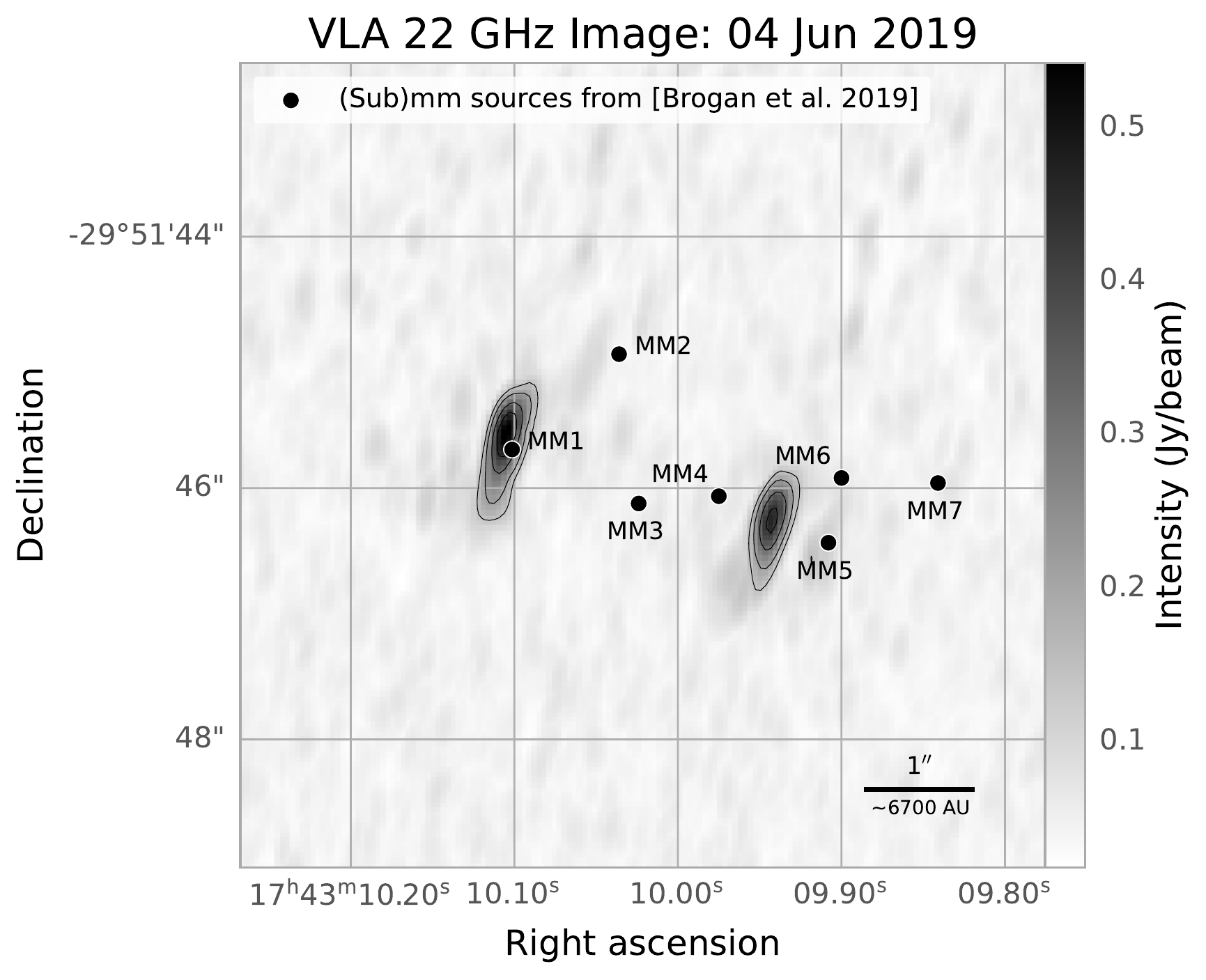} \\
(a)  & (b)  \\[6pt]
\end{tabular}
\caption{22 GHz H$_2$O maser emission detected in \object{G358.93$-$0.03} with the VLA during (a) epoch I and (b) epoch II. 
\label{fig:22img}}
\end{figure*}

The spatial distribution of the detected H$_2$O masers is found to differ significantly from the methanol masers detected with the VLA during the same observing sessions and described in \cite{Bayandina2022}.
The water maser emission covers a larger area  of about~3$^{\prime\prime}$ (Figs. \ref{fig:22compare} and \ref{fig:22img}), while the methanol emission was detected in a region of $\sim$0.2$^{\prime\prime}$ in the vicinity of MM1 only.

\begin{table*}
\caption{22 GHz H$_2$O maser parameters.}
\label{tab:T22GHZ}
\centering  
\begin{tabular}{cccccc}
\hline\hline   
Component &
RA(J2000) & Dec.(J2000)  &
Integrated flux &
Peak flux &
V$_{LSR}$ \\ &
($^h$~$^m$~$^s$) & ($^\circ$~$\arcmin$~$\arcsec$)  &
(mJy) & 
(mJy/beam) &
(km~s$^{-1}$) \\
\hline
I-1 & 17:43:10.1037$\pm$0.0001 & -29:51:45.609$\pm$0.038 & 235$\pm$22  & 222$\pm$9.1  & -20.13\\
& 17:43:10.1018$\pm$0.0002 & -29:51:45.579$\pm$0.039 & 175$\pm$22  & 201$\pm$10   & -20.08\\
& 17:43:10.1026$\pm$0.0002 & -29:51:45.585$\pm$0.041 & 231$\pm$23  & 224$\pm$9.4  & -20.02\\ 
\hline                                   
\end{tabular}
\tablefoot{
(1) Table \ref{tab:T22GHZ} is published in its entirety in machine-readable format. A portion is shown here for guidance regarding its form and content. \\ (2) The positional shifts of $\Delta$RA=-0.02, $\Delta$Dec.=-0.02 (epochs I and II) were introduced to the data to prepare the figures (see Sect. \ref{sec:obs}).}
\end{table*}

During epoch I, the spatial distribution of the H$_2$O masers consists of four spatial components (Fig. \ref{fig:22compare}-\ref{fig:22img}a): Component I-1 is found in the vicinity of MM1 and hosts maser spots with a velocity range from $\sim -17.5$ to $-20$ km~s$^{-1}$. Component I-2 is the brightest water maser in the region. It is located $\sim$1$^{\prime\prime}$ to the NW of MM1 and is spatially associated with the continuum source MM2. Maser spots detected in Component I-2 have the same velocities as those of Component I-1. Components I-3 and I-4 are found to the SW of MM1 and are associated with the continuum sources MM4 and MM5, respectively, detected in \cite{Brogan19}. Component I-3 consists of the maser spots with velocities of $\sim-17$ km~s$^{-1}$, and Component I-4 shows velocities of $\sim-20$ km~s$^{-1}$.
Thus, during epoch I water masers are observed towards all the strongest millimetre sources, except MM3, which is the only source with associated continuum emission. 

In contrast, only two H$_2$O maser components are detected during VLA epoch II (Fig. \ref{fig:22compare}-\ref{fig:22img}b): Component II-1 is found at the same position as Component I-1, close to the line-rich source MM1, but the spectrum of Component II-1 has a wider velocity range and spans from $\sim-13.5$ to $-20.5$ km~s$^{-1}$. No emission is detected at the position of Component I-2. Component II-3-4 is located between 
the positions of Components I-3 and I-4 and, consequently, between the continuum sources with which these water masers were associated, MM4 and MM5. The component consists of maser spots with velocities of $\sim-21$ to $-22$ km~s$^{-1}$. No features with these velocities were detected during epoch I.

\subsubsection{Water masers associated with MM1}
The bursting source MM1 showed  water maser emission during both VLA epochs, though the distribution of the maser spots changed significantly between the observations. 

\begin{figure*}[ht]
\centering
\begin{tabular}{cc}
  \includegraphics[width=65mm]{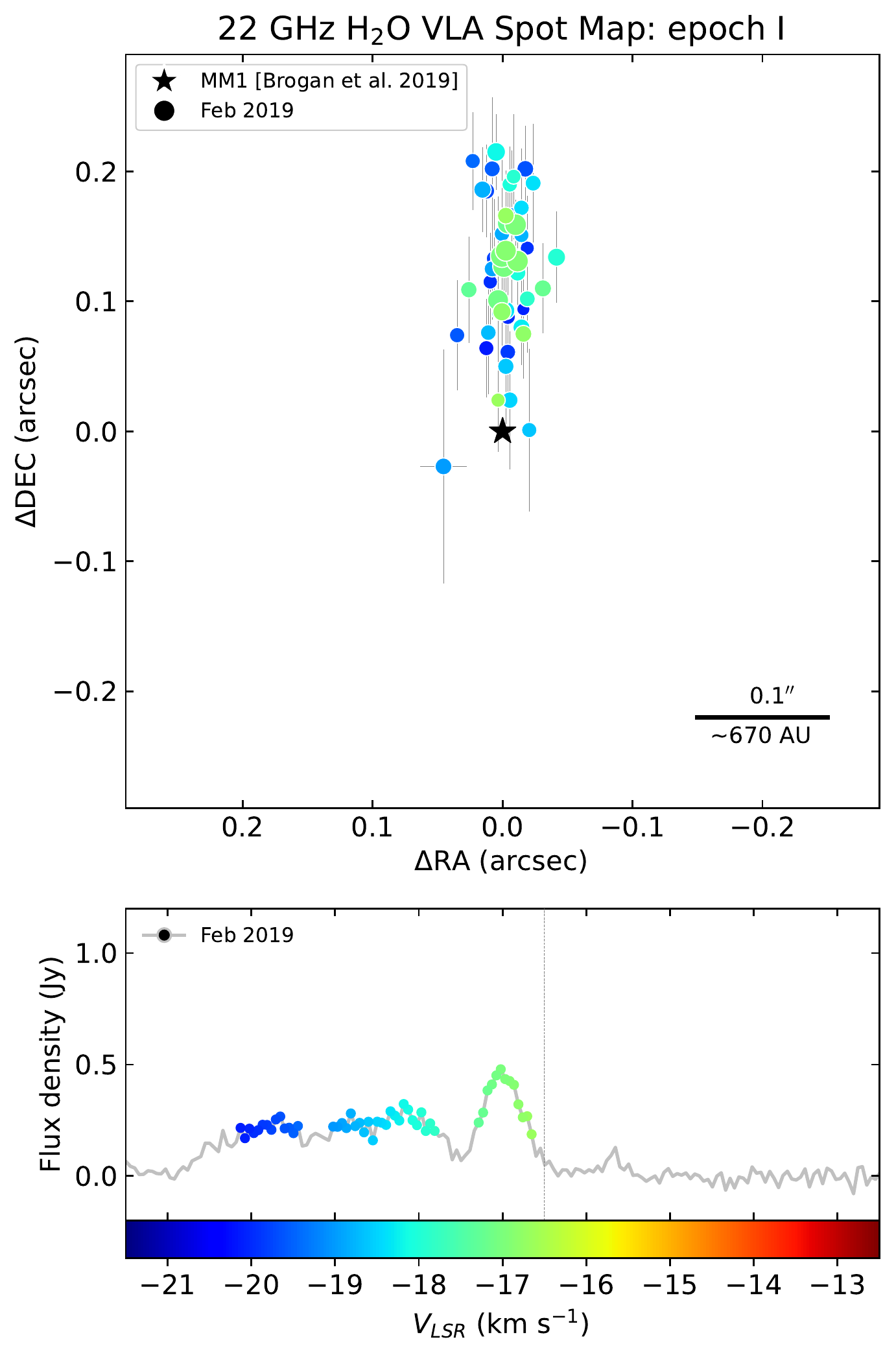} &   
  \includegraphics[width=65mm]{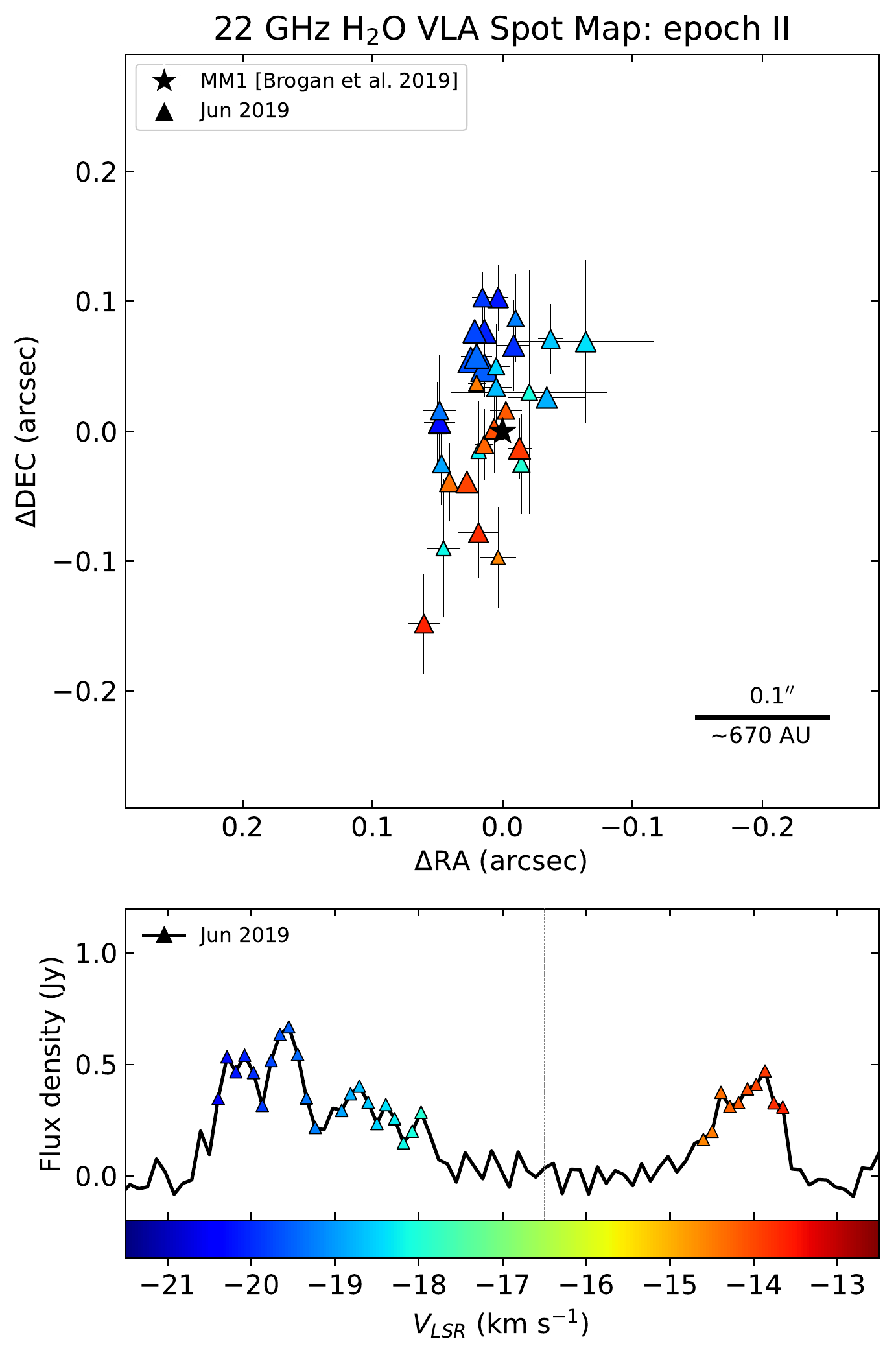} \\
(a)  & (b)  \\[6pt]
\end{tabular}
\caption{22 GHz water maser spot maps and spectra of Component I-1 from epoch I and Component II-1 from epoch II. The plots are colour-coded by radial velocity (see the colour bar for the colour scale). The markers on the spectra correspond to the maser spots on the map. The diameter of each spot is proportional to the flux. The error bars indicate the position fitting errors from Table~\ref{tab:T22GHZ}. Positional offsets are relative to the MM1 position (star marker) from \cite{Brogan19}. The dashed line in the spectrum panels marks the centre velocity of MM1 \cite{Brogan19}. \label{fig:spotsM1}}
\end{figure*}

During epoch I, the maser spots composing Component I-1 spread from MM1 to the north (Fig. \ref{fig:spotsM1}). The cluster has a linear structure with  a length of $\sim$0.2$^{\prime\prime}$ and a width of only $\sim$0.05$^{\prime\prime}$.
No clear velocity gradient is found. 
The distribution of the maser spots resembles the pre-H$_2$O maser flare state detected in the source with the VLA B-configuration on April 4, 2019, and reported in \citet{Chen20b}.
However, while the general spatial distribution is preserved, the velocity range of the water maser emission detected in  \citet{Chen20b} was broader (V$_{LSR}$ from $\sim$$-$16 to $\sim$$-$23 km~s$^{-1}$).

In contrast to epoch I, the 22 GHz water maser cluster II-1 is distributed on both sides of MM1 and elongated 0.2$^{\prime\prime}$ in the NW-SE direction (Fig. \ref{fig:spotsM1}). 
It shows a  velocity gradient similar to the one found for the methanol masers in \cite{Bayandina2022}:  blueshifted  maser spots are displaced to the north and redshifted spots to the south. \\

\subsubsection{Water masers associated with MM2, MM4, and MM5}
\begin{figure*}[ht]
\centering
\begin{tabular}{ccc}
  \includegraphics[width=50mm]{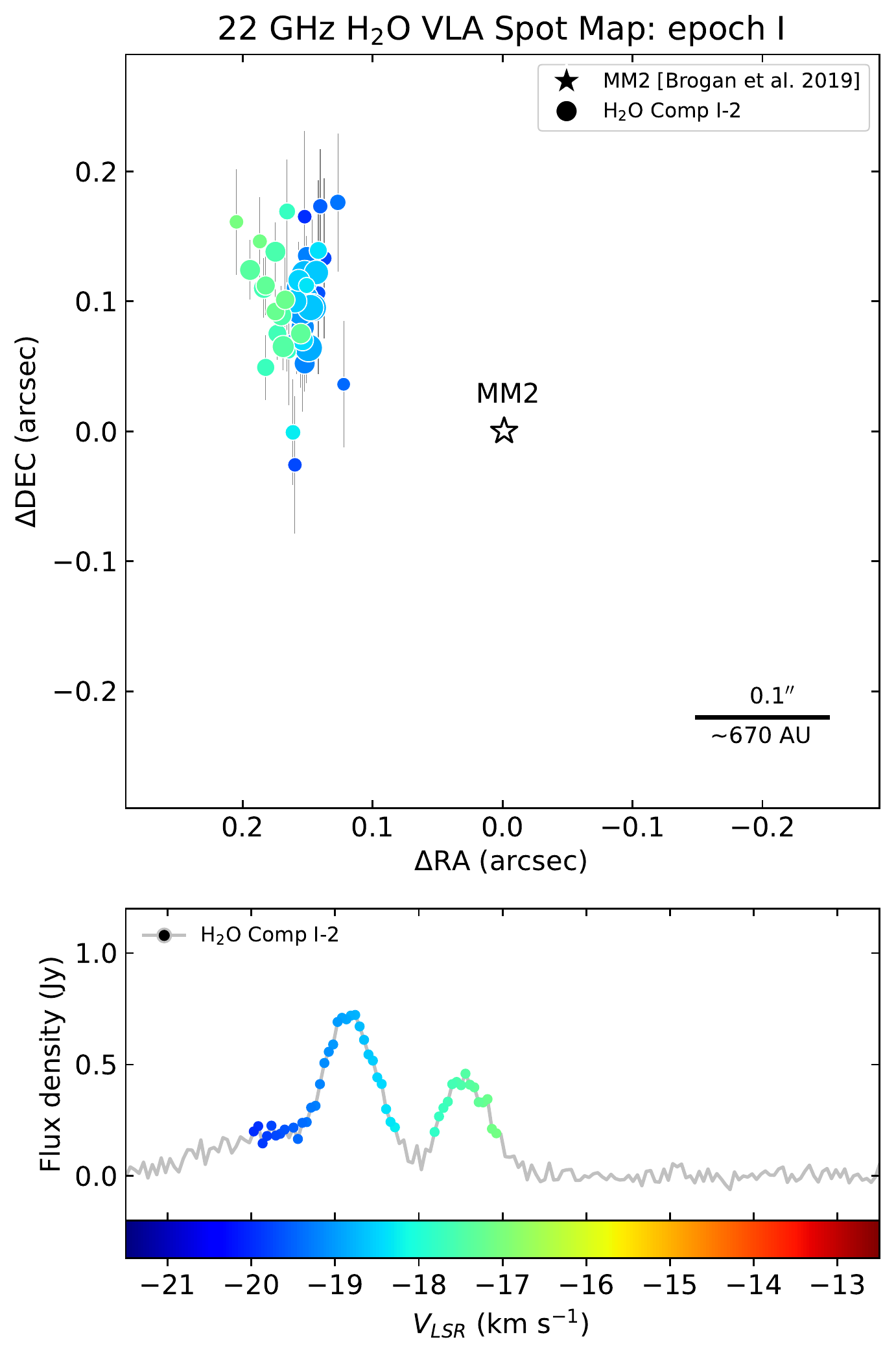} &   
  \includegraphics[width=50mm]{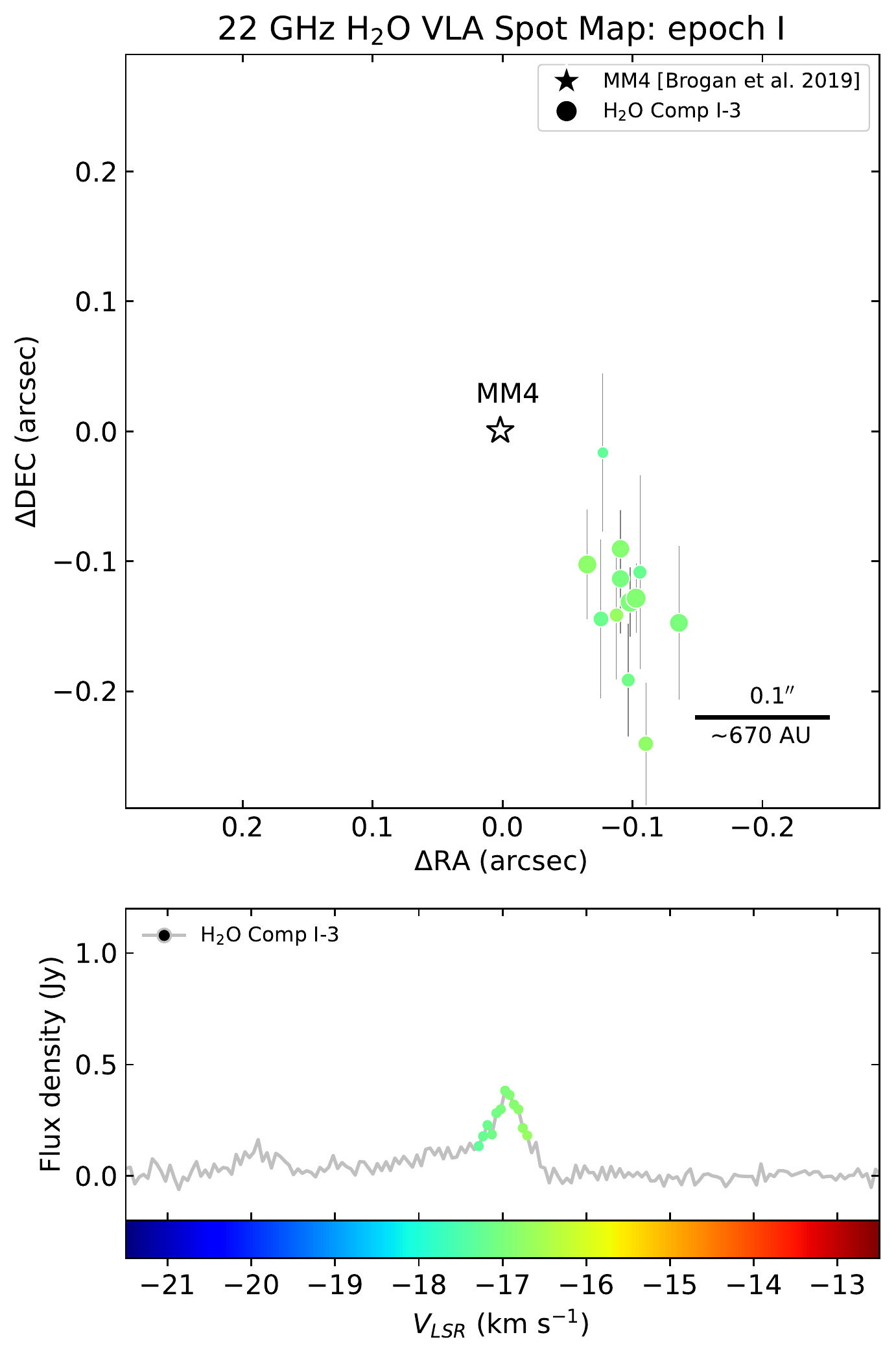} & 
  \includegraphics[width=50mm]{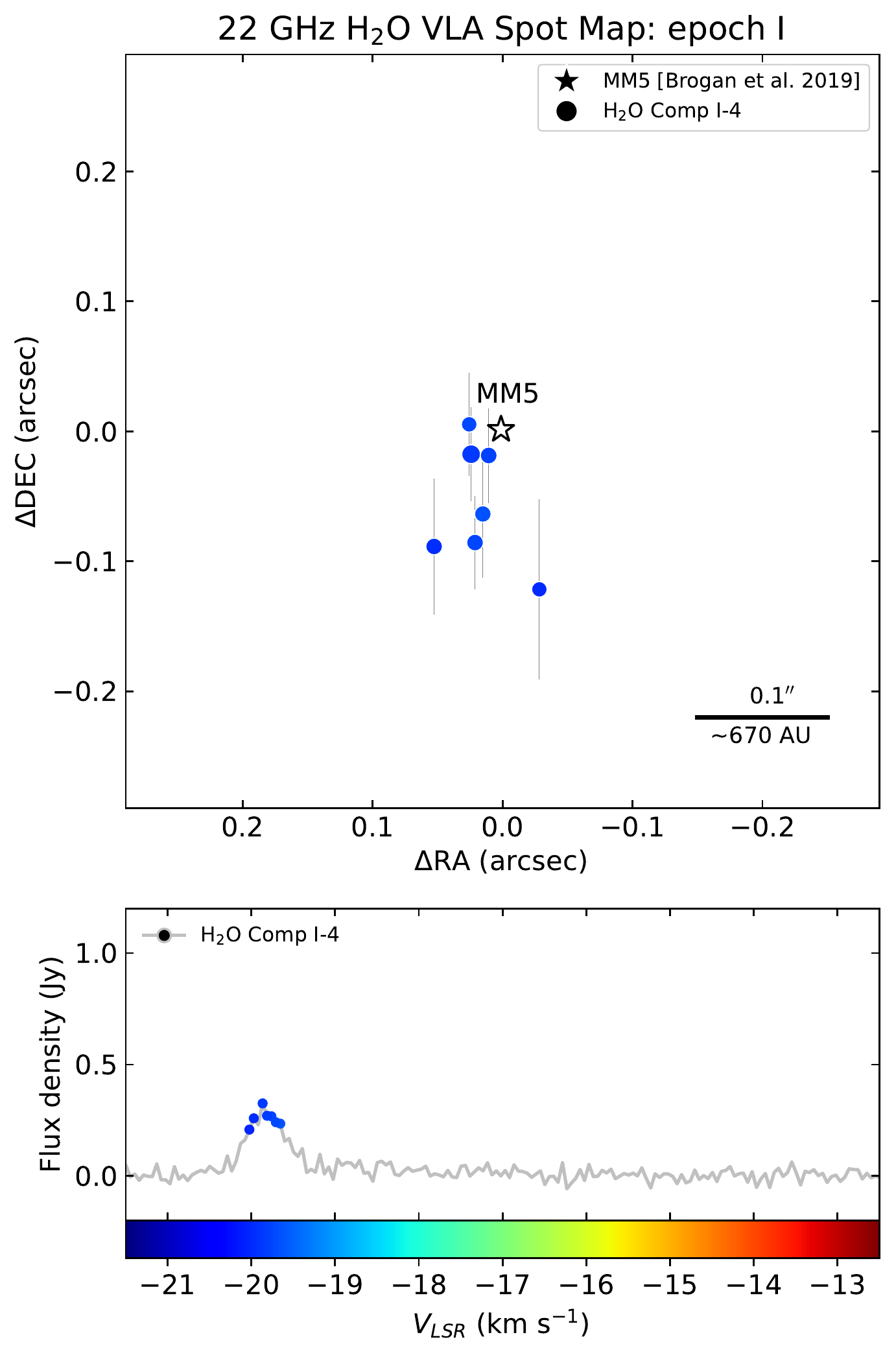}\\
(a)  & (b) & (c) \\[6pt]
\end{tabular}
\caption{22 GHz water maser spot maps and spectra of Components (a) I-2, (b) I-3, and (c) I-4 from epoch I. The plots are colour-coded by radial velocity (see the colour bars for the colour scale). The markers on the spectra correspond to the maser spots on the map. The diameter of each spot is proportional to the flux. 
The positional offsets are relative to the (a) MM2, (b) MM4, and (c) MM5 (star markers) positions from \cite{Brogan19}. \label{fig:spotsM25}}
\end{figure*}

\begin{figure}[ht]
\centering
\includegraphics[width=0.30\textwidth]{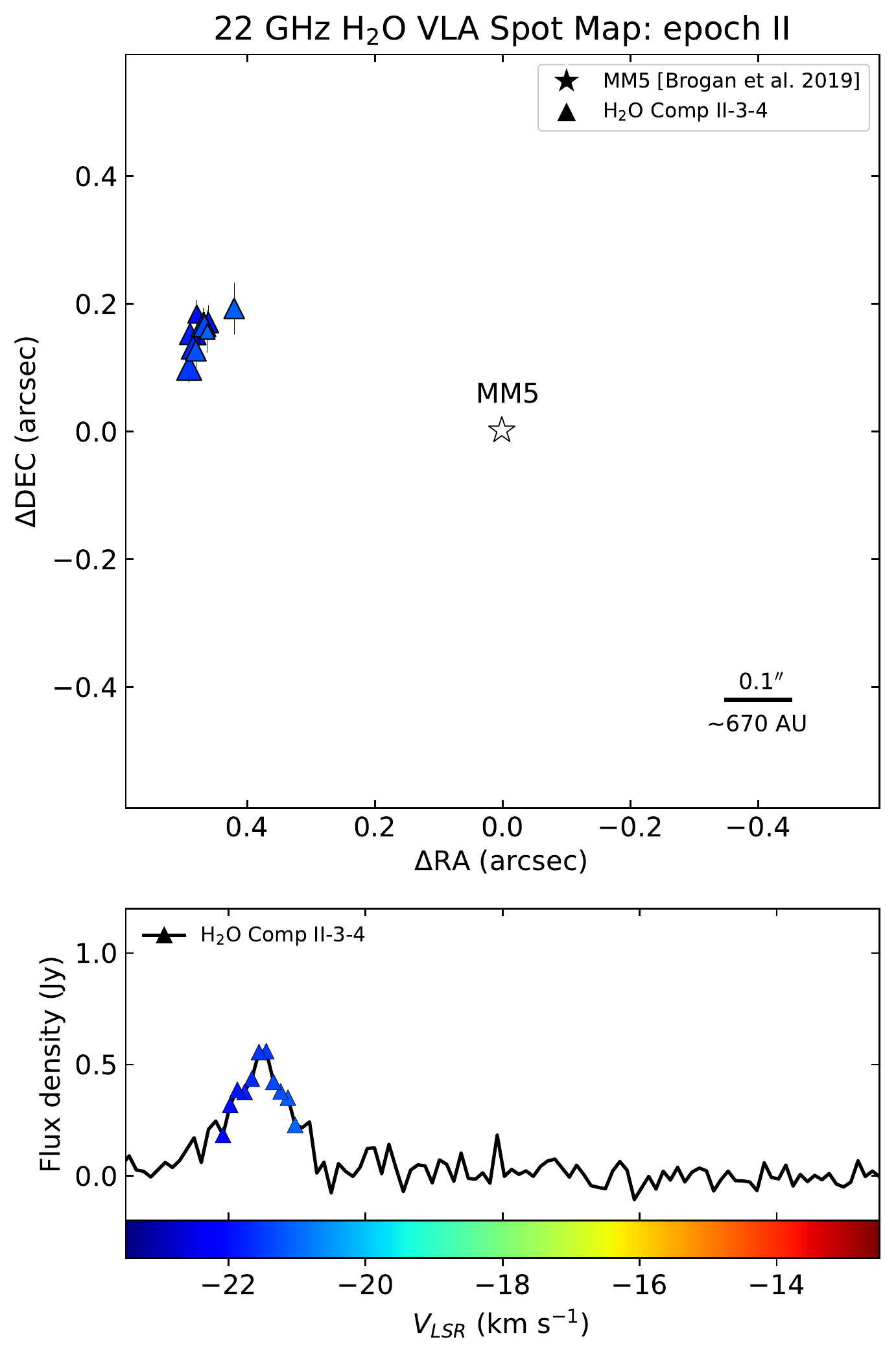}
\caption{22 GHz water maser spot map and spectrum of Component II-3-4 from epoch II. The plot is colour-coded by radial velocity (see the colour bar for the colour scale). The markers on the spectrum correspond to the maser spots on the map. The diameter of each spot is proportional to the flux. 
The positional offsets are relative to the MM5 position (star marker) from \cite{Brogan19}. \label{fig:spotsII34}}
\end{figure}

Water masers, spatially associated with the continuum sources MM2 and MM4, are found during VLA epoch I only (Fig.~\ref{fig:spotsM25}). The water masers in the vicinity of MM5 are found during both epochs (Figs. \ref{fig:spotsM25} and \ref{fig:spotsII34}) but show significant spatial shift. 

The H$_2$O masers associated with MM2, Component I-2, are located  $\sim$0.15$\arcsec$ to the NE of the continuum source. The maser spots correspond to two spectral features at  velocities of $-$18.8 and $-$17.5 km~s$^{-1}$ and form two close, but clearly spatially separated, clusters. The cluster at velocities of about $-$19 km~s$^{-1}$ is located closer to MM2 and is linearly elongated in the N-S direction. The other cluster at about $-$17 km~s$^{-1}$ is separated more from MM2 and shows a linearly elongated structure and an orientation similar to the previous one.

Component I-3 is located  $\sim$0.1$\arcsec$  SW  of MM4. The cluster is linearly elongated with a size of $\sim$0.2$\arcsec$.

The water maser emission of Component I-4 is found  close to MM5 and linearly elongated $\sim$0.1$\arcsec$ to the south. Similar to Component I-1, it is closely associated with its continuum source. Component I-4 is the least crowded and consists of fewer than a dozen maser spots. 

Component II-3-4 is the only water maser besides Component II-1 detected during VLA epoch II. Its peak velocity is about -21.5 km~s$^{-1}$. The component is located at about the same separation of $\sim$0.48$\arcsec$ from the positions of MM4 and MM5, but based on the maser spots' velocity and cluster orientation, it seems to be associated with MM5. The cluster appears as a bow-shaped structure  elongated in the NW-SW direction, with a curvature radius that corresponds approximately to the direction towards MM5.

\section{Discussion} \label{sec:discussion}

\subsection{Ejection activity of the accretion burst source MM1}

\begin{figure*}[ht]
\centering
\begin{tabular}{ccc}
  \includegraphics[width=80mm]{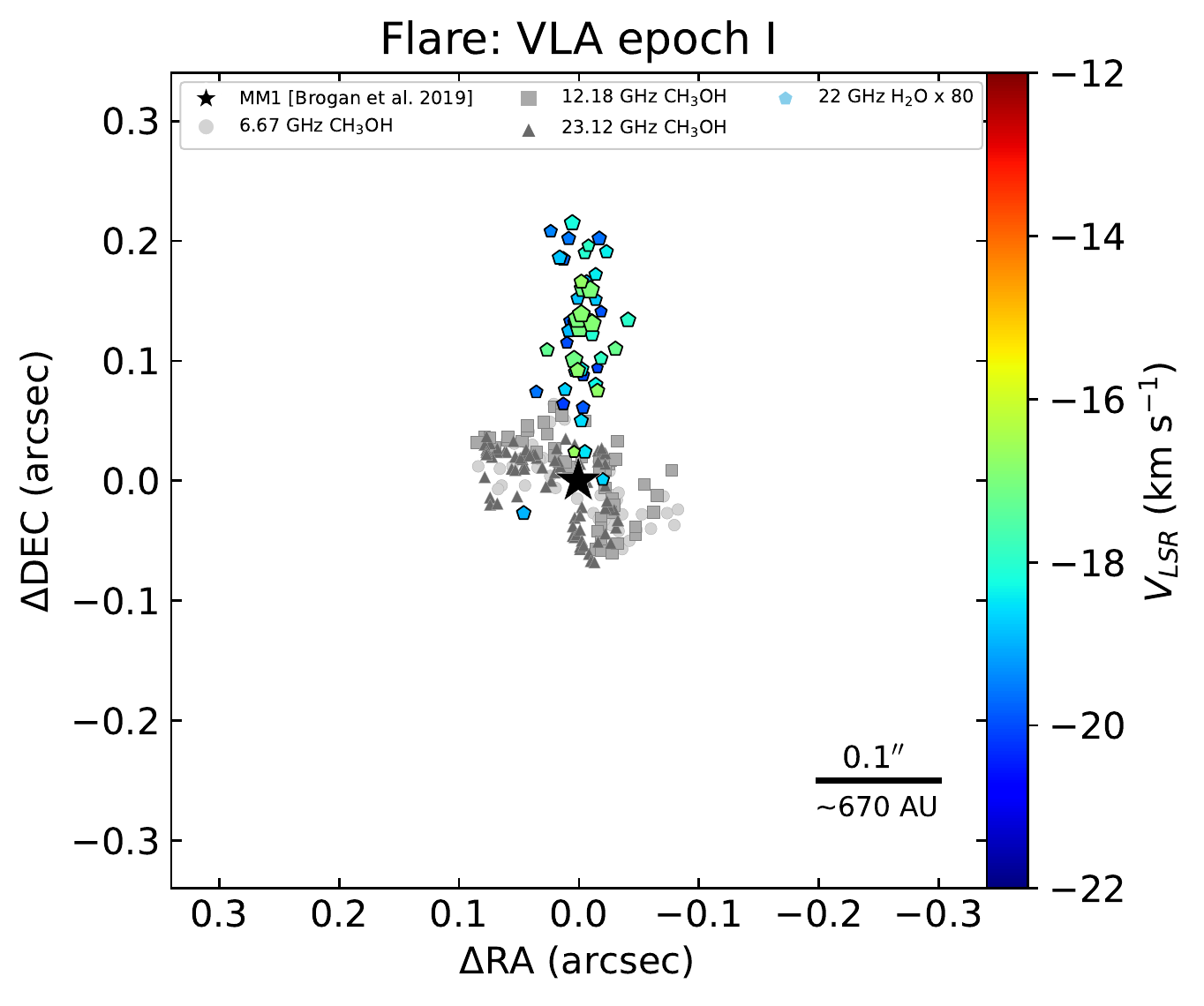} &   
  \includegraphics[width=80mm]{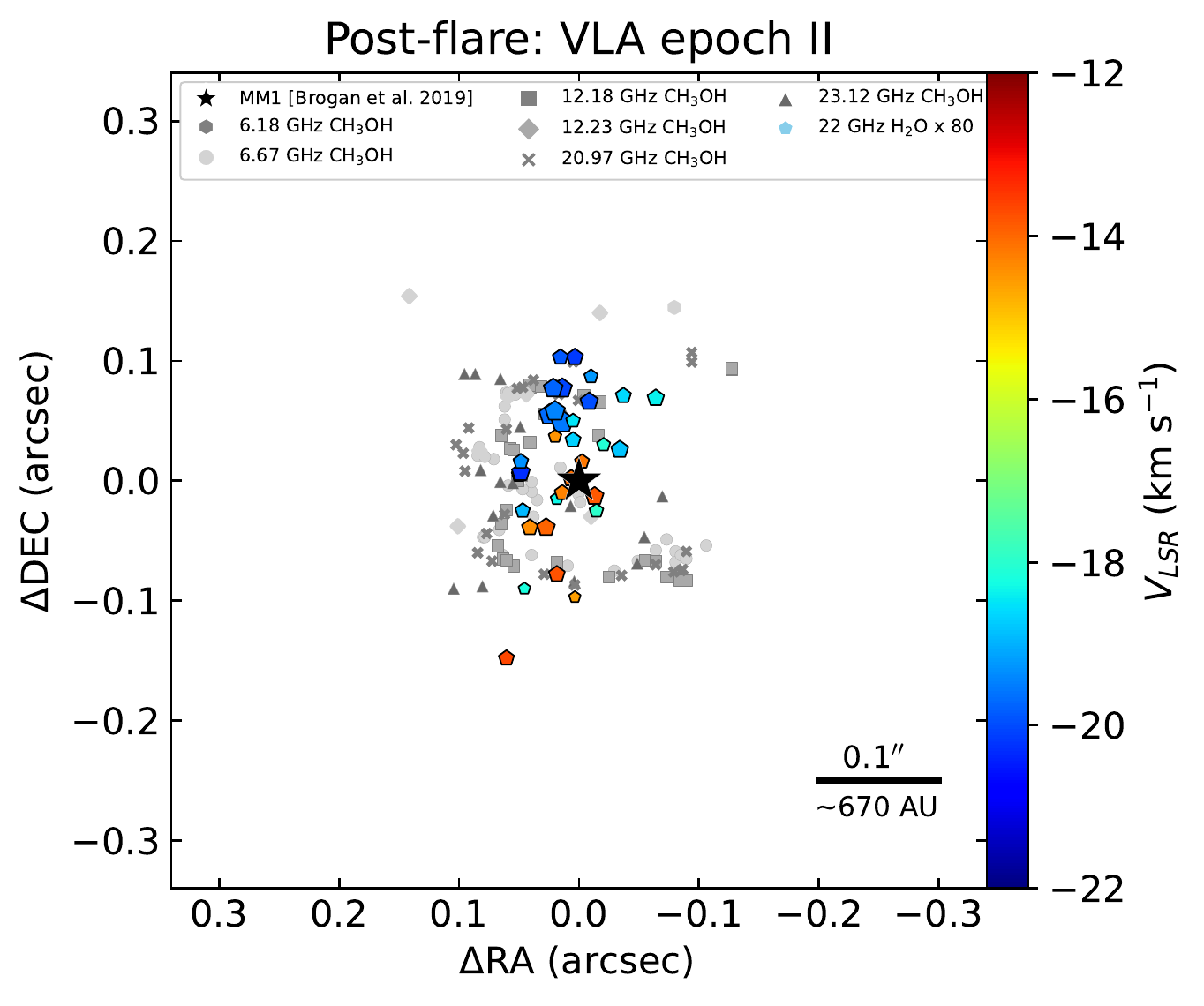} \\
(a)  & (b) \\[6pt]
\end{tabular}
\caption{Comparison of spatial distribution of the methanol from \cite{Bayandina2022} and water maser emission detected in \object{G358.93$-$0.03} during the (a) flare (VLA epoch I) and  (b) post-flare (VLA epoch II) epochs.
Positional offsets are relative to the MM1 position from \cite{Brogan19} (star marker).
\label{fig:compare2}}
\end{figure*}

The narrow, elongated spatial structure of the water maser clusters detected in the vicinity of MM1 (Components I-1 and II-1; Fig. \ref{fig:spotsM1}) as well as their location relative to the accretion disk traced by the methanol masers (Fig.
\ref{fig:compare2}) suggest their association with a jet or outflow powered by MM1. 
Considering the low flux density ($\sim$1~Jy) of the water maser emission and the size of the maser clusters ($\sim$0.2$\arcsec$) detected during two VLA epochs, the jet seems to be at an early phase of activity. However, we must note that H$_2$O masers are not good tracers of jet size as they trace only the densest portion of it. 

One feature of the source, which has complicated the study of the ejection caused by the accretion burst, is the orientation of the source in the plane of the sky. A face-on accretion disk around MM1 was detected by the methanol maser observations  \citep{Burns20,Chen20b,Bayandina2022}. The orientation of the disk allowed us to trace the propagation of the burst heatwave and to penetrate the fine structure of the disk, which turned out to be composed of spiral arm structures \citep{Chen20b,Bayandina2022}. However, assuming the axis of the outflow to be perpendicular to the disk, the outflow projection effect to the plane of the sky means that the blueshifted and redshifted outflows are largely coincident. 
Our VLA H$_2$O maser results (especially epoch II; see Fig. \ref{fig:spotsM1}), which show partial spatial overlap between the blueshifted and redshifted maser features, support our suggested
orientation of the outflow along the line-of-sight to the source.

\begin{figure}[ht]
\centering
\includegraphics[width=0.5\textwidth]{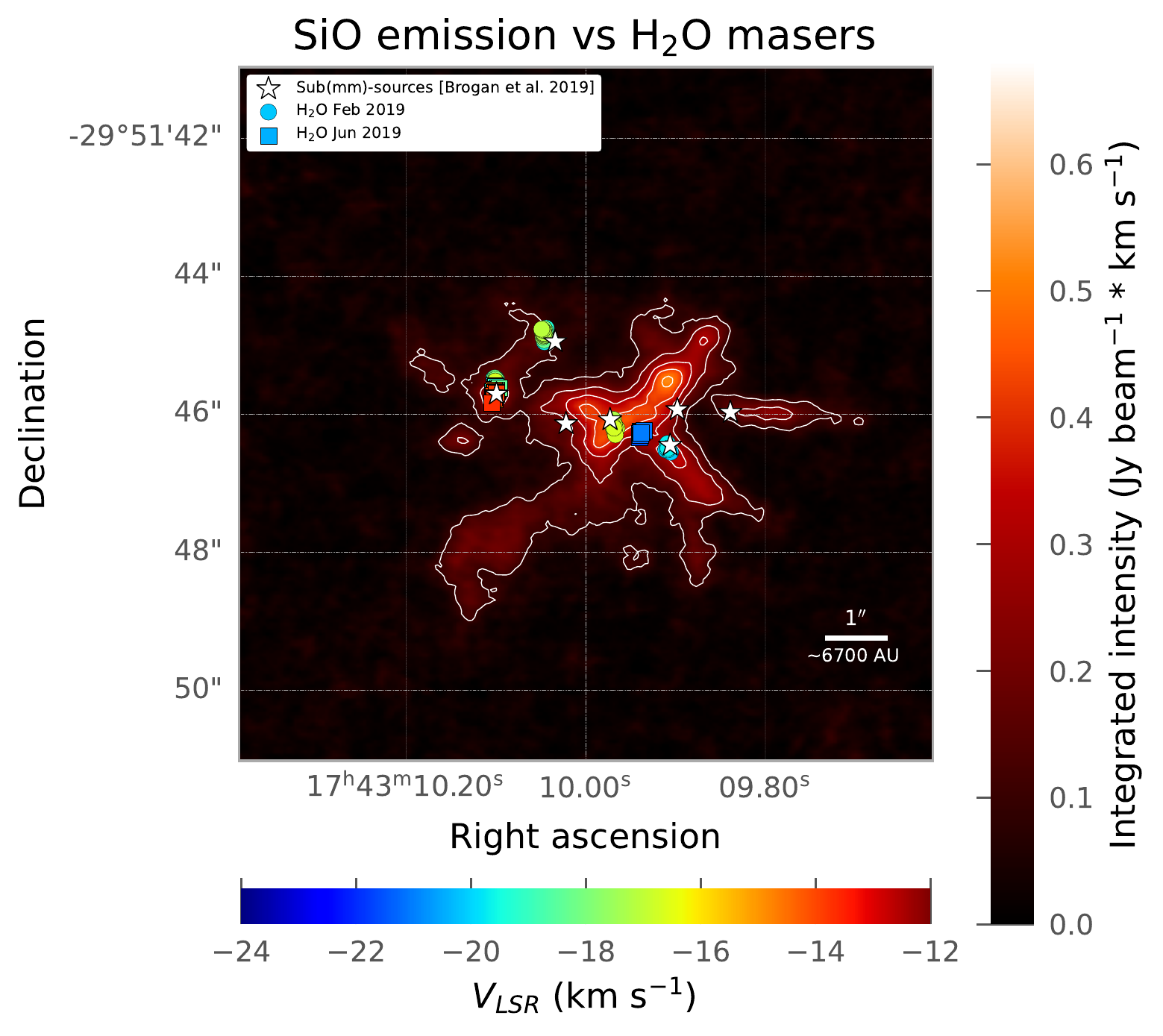}
\caption{22 GHz water maser spot map overplotted on the moment 0 map (integrated over the velocity span of -40 to 3.2 km s$^{-1}$) of the SiO J=5-4 emission obtained with ALMA. The contour levels are
[1, 2, 3, 4, 5] $\times$ 0.1~Jy beam$^{-1}$ km s$^{-1}$. 
\label{fig:sio}}
\end{figure}

Since both H$_2$O masers and SiO emission are typical tracers of outflow activity, we compared our VLA observations with  archive ALMA data (Project: 2019.1.00768.S) of SiO J=5-4 emission in the region (Fig. \ref{fig:sio}).
The SiO image was obtained with the angular resolution of $\sim$0.3$\arcsec$, combining two ALMA datasets from 2019 (October 1 and 7) and one from 2021 (April 1). 
The ALMA SiO emission associated with MM1 shows a compact structure with a size of $\sim$0.7\arcsec\ (corresponding to $\sim$4~700 AU at the distance of 6.75 kpc)  in all channels (see Fig. \ref{fig:sio}).
A more detailed analysis of the properties of the outflows in the region based on ALMA data will be presented in Brogan et al. in prep.

The evolution of the water maser structure between the two VLA observation epochs (Fig. \ref{fig:spotsM1}) was assumed to be excited by an enhanced radiation field caused by the accretion event, and not a mechanical matter transfer of the circumstellar medium.
The southern part of the H$_2$O maser emission with the redshifted velocities has a size of $\sim$0.1$\arcsec$ and was detected  during VLA epoch II only. If we assume the presence of a physical matter movement that travelled $\sim$0.1$\arcsec$ (or $\sim$670 AU at the distance of 6.75 kpc) in 99 days between VLA epochs I and II, it must have had a speed of 0.01c. Such speeds are too high for the survival of molecules, including H$_2$O.
Therefore, it is more likely that the water masers detected during the first  observation epoch dimmed,
and the new masers, detected during the second epoch, were excited by the heating of the medium by the arrived radiation with proper spectral properties. Such a scenario was proposed for \object{NGC~6334~I} \citep{Sobolev19,Brogan18}.

\subsection{Large-scale effect of the accretion burst told by H$_2$O masers}

Our VLA images have revealed drastic changes in the distribution of the water masers in the whole region and not only around the bursting source MM1. Of the three clusters detected during the first VLA epoch and associated with MM2, MM4, and MM5, only one associated with MM5 remained after the burst (Fig.~\ref{fig:22img}).

The arc-shaped morphology of Component I-2 (Fig~\ref{fig:spotsM25}) suggests its association with  outflow activity in the vicinity of MM2 \citep[e.g.][]{Burns2017}. Under this assumption, the water masers trace bow-shaped shocks  expanding from the central source, MM2. Two separate `layers' of the shock are highlighted by the water masers. The layers overlap spatially (within uncertainties) but are clearly separated in terms of velocity. The brighter layer ($\sim$0.8 Jy) has a bluer V$_{LSR}$ of $\sim$-19 km~s$^{-1}$, and the fainter layer ($\sim$0.5 Jy) has a redder V$_{LSR}$ of $\sim$-17.5 km~s$^{-1}$. We note that the MM2 region shows weak SiO emission (Fig. \ref{fig:sio}). 
The component was the brightest water maser in the region during epoch I, but it became undetectable during epoch II. The change in the flux density can be attributed to the high variability of the water maser, which, coupled with the described morphology of the component, suggests that MM2 is the driving source of a young and active jet or outflow.
On the other hand,  the spectral feature at V$_{LSR}$ = -17.42 km~s$^{-1}$ of Component I-2 precisely coincides in terms of velocity with the peak of the water maser flare (see Fig.~\ref{fig:22compare}). Based on this curious coincidence, we can speculate that the strongest water maser flare, detected on April 22, 2019, may have been caused by an increase in the brightness of the water masers in the vicinity of MM2. Unfortunately, the lack of knowledge about the spatial distribution of water masers in the source at the time of the flare prevents us from verifying this assumption. 

A bright and extended peak of the SiO emission is found in the vicinity of MM4. While the  SiO emission suggests the presence of  outflow activity associated with the continuum source, the  water maser Component I-3 is faint ($\sim$0.5 Jy) and detected only during epoch I. The elongated structure of the water maser might be associated with a jet or outflow. The SiO and water maser emission trace shock activity on different sides of MM4: the axis of the SiO outflow is oriented to the NE, while the water maser jet or outflow  axis points to the SW.
Based on the SiO image, MM4 seems to be one of the most active outflow sources in the region.

Apart from MM1, the only point source that showed water maser activity during both VLA epochs is MM5.
The ALMA SiO emission  associated with MM5 shows a NE-SW elongation, and we infer this to be the axis of the bipolar outflow driven by the source.
While the water maser detected during epoch I (Component I-4) is faint and spatially scattered (though it is clearly elongated to the south), the second epoch maser (Component II-3-4) has  an arc-shaped morphology and, similarly to Component I-2, may trace bow-shaped shocks from MM5. 
The post-flare water masers are probably tracing a bow shock along the jet, directed perpendicular to the pre-flare elongation of the water masers, which therefore could arise close to the disk surface.

Water masers are well known to be variable, and, given the separation of MM1 from the three water maser components associated with MM2, MM4, and MM5, it is quite possible that their behaviour is unrelated to the burst event in MM1. Unfortunately, because of the low flux density and spectral features blending, monitoring of the water maser emission in the source is challenging. No long-term monitoring information on the H$_2$O maser emission variability is available for the source, which limits our analysis of this option.

On the other hand, the propagation speed of a heatwave of an accretion burst is different in different directions: in the plane of an accretion disk, the wave is slowed  by dense structures (as shown in \citealt{Burns20,Bayandina2022}), but in  directions where the density of matter is very low, the heatwave speed is supposed to be close to the speed of light in a vacuum. The size of the accretion disk around \object{G358.93$-$0.03-MM1} is much smaller than the distance to the water masers in the region (radii of $\sim$0.1$\arcsec$ vs. $\sim$2$\arcsec$ or $\sim$6~750 vs. $\sim$13~500 AU, respectively); however, the thermal wave mainly moves through  `empty' space on its way to the water masers.
Of particular interest is the fact that, based on this hypothesis, the H$_2$O maser flare was predicted  to start in April (Fig.~\ref{fig:timeline}; A.M.~Sobolev, private communication). 
The onset of the accretion burst was assumed to coincide with the methanol maser flare (mid-January 2019; see  e.g. Fig. 1 in \citealt{Bayandina2022}), and, given the H$_2$O masers' separation from MM1 of $\sim$2$\arcsec$ (or $\sim$13~500 AU; see Fig. \ref{fig:22compare}) and the source distance of 6.75~kpc \citep{Reid09}, the light travel time between MM1 and the water maser positions was supposed to be  $\sim$3 months.

Under such an assumption, the far-infrared emission from MM1 has reached the disk or envelope of other YSOs in the region and has increased the gas temperature and inner turbulence, thus disrupting  the velocity coherence for water maser emission (if arising close to the disk surface). 
Assuming a disk-wind origin for the water masers, a general interpretation might be that the increased temperature and turbulence of the envelope or disk owing to the accretion burst radiation has quenched the (slower) water masers near the YSOs, allowing only for those that could be associated with (faster) shocks at greater distances along the jet or outflow. 
The detection of several water maser flares (Fig.~\ref{fig:timeline}) may be due to the fact that the thermal wave is decelerated in different dense formations, its propagation is not rectilinear (the direction changes), and the heating of the water maser region by the radiation of an accretion burst can occur several times.

We note that the behaviour of the H$_2$O masers in previously known sources of MYSO accretion bursts was proposed to be explained by the propagation of the burst radiation along the outflow cavities (see \citealt{Brogan18} for \object{NGC6334I-MM1}  and  \citealt{Hirota2021} for \object{S255~NIRS~3}). 
In the case of \object{S255 NIRS 3}, even the presence of H$_2$O masers pumped by radiation (theoretically predicted by \citealt{Gray16}) has been suggested \citep{Hirota2021}.

\section{Conclusions}

Two epochs of VLA imaging of the 22 GHz water maser emission and continuum in the C, Ku, and K bands were performed for the  massive  star-forming  region \object{G358.93$-$0.03}. 

   \begin{enumerate}
      \item A drastic change in the distribution of the 22 GHz water masers in the region is found: four spatial maser components are detected during epoch I, and only two during epoch II. 
      \item The 22 GHz water maser associated with the source of the accretion burst, MM1, traces a jet (outflow) oriented towards the observer with partly overlapping lobes in the plane of the sky. 
      \item During epoch I, only the blueshifted lobe of the H$_2$O jet in MM1 is detected. 
      \item During epoch II, both blueshifted and redshifted lobes of the MM1 jet are detected, indicating the change in the source environment properties to the ones favourable for H$_2$O maser emission.
      \item The extinction of the H$_2$O masers associated with the MM1 neighbouring continuum sources during epoch II is assumed to indicate a large-scale effect of an accretion event on the host region. 
       \item Continuum emission associated with the hot core \object{G358.93$-$0.03-MM3} is detected in all observed frequency bands. 
   \end{enumerate}
   
   The presented  VLA study of the 22 GHz water masers in the accretion burst source \object{G358.93$-$0.03} illustrated the great importance of observing ejection tracers in the aftermath of an accretion burst, since they not only probe the environment of the central source, but can also reveal possible large-scale burst effects on the entire host region. 

\begin{acknowledgements}
The Ibaraki 6.7-GHz Methanol Maser Monitor (iMet) program is partially supported by the Inter-university collaborative project ”Japanese VLBI Network (JVN)” of NAOJ and JSPS KAKENHI Grant Numbers JP24340034, JP21H01120, and JP21H00032 (YY). 

The National Radio Astronomy Observatory is a facility of the National Science Foundation operated under
cooperative agreement by Associated Universities, Inc. 
This paper makes use of the following ALMA data: ADS/JAO.ALMA\#2019.1.00768.S. ALMA is a partnership of ESO (representing its member states), NSF (USA) and NINS (Japan), together with NRC (Canada), MOST and ASIAA (Taiwan), and KASI (Republic of Korea), in cooperation with the Republic of Chile. The Joint ALMA Observatory is operated by ESO, AUI/NRAO and NAOJ. In addition, publications from NA authors must include the standard NRAO acknowledgement: The National Radio Astronomy Observatory is a facility of the National Science Foundation operated under cooperative agreement by Associated Universities, Inc. AMS was supported by the Russian Science Foundation grant 18-12-00193. A.C.G. acknowledges support by PRIN-INAF-MAIN-STREAM 2017 "Protoplanetary disks seen through the eyes of new-generation instruments" and by PRIN-INAF 2019 "Spectroscopically tracing the disk dispersal evolution (STRADE)".
\end{acknowledgements}

%
%

 \bibliographystyle{aa}
 \scriptsize{ 
  \bibliography{bibliography}}

\end{document}